\documentclass[journal,onecolumn,final]{IEEEtran}

\usepackage{cite}
\usepackage[latin9]{inputenc}
\usepackage{amsthm}
\usepackage{amsmath}
\usepackage{amssymb}
\usepackage{cite}
\usepackage{amssymb,amsmath}
\usepackage{enumerate}
\usepackage{centernot}
\usepackage{multicol,comment}
\usepackage{xspace}
\usepackage{enumitem}
\usepackage{environ}
\usepackage{times}
\usepackage{float}
\usepackage{algorithmicx}
\usepackage{algpseudocode}
\usepackage{lscape}
\usepackage{pdflscape}
\usepackage{wrapfig}
\usepackage{rotating}
\usepackage{epstopdf}
\usepackage{setspace}

\usepackage{algorithm}

\usepackage{graphicx,subcaption}

\usepackage[unicode,pdfstartview=FitH]{hyperref}
\makeatletter

\newcommand{\lyxmathsym}[1]{\ifmmode\begingroup\def\b@ld{bold}
  \text{\ifx\math@version\b@ld\bfseries\fi#1}\endgroup\else#1\fi}

\theoremstyle{plain}

\theoremstyle{plain}

\usepackage{epsfig}\usepackage{cite}

\makeatother

\providecommand{\lemmaname}{Lemma}
\providecommand{\theoremname}{Theorem}

\begin{document}

\title{Image Block Loss Restoration Using Sparsity Pattern as Side Information}

\author{Hossein Hosseini,~\IEEEmembership{Student Member,~IEEE}, Ali Goli, Neda Barzegar Marvasti, Masoumeh Azghani and Farokh Marvasti,~\IEEEmembership{Senior Member,~IEEE}\\
Advanced Communication Research Institute (ACRI)\\
Department of Electrical Engineering, Sharif University of Technology, Tehran, Iran\\
hosseinh@uw.edu, goli@mie.utoronto.ca, neda.barzegarmarvasti@boun.edu.tr,\\
 azghani@ee.sharif.edu, marvasti@sharif.edu }
\maketitle

\begin{abstract}
In this paper, we propose a method for image block loss restoration based on the notion of sparse representation. We use the sparsity pattern as side information to efficiently restore block losses by iteratively imposing the constraints of spatial and transform domains on the corrupted image. Two novel features, including a pre-interpolation and a criterion for stopping the iterations, are proposed to improve the performance. Also, to deal with practical applications, we develop a technique to transmit the side information along with the image. In this technique, we first compress the side information and then embed its LDPC coded version in the least significant bits of the image pixels. This technique ensures the error-free transmission of the side information, while causing only a small perturbation on the transmitted image. Mathematical analysis and extensive simulations are performed to validate the method and investigate the efficiency of the proposed techniques. The results verify that the proposed method outperforms its counterparts for image block loss restoration.

Index Terms\textemdash{}Image Restoration, Block Loss, Image Sparsity, Sparsity Pattern, Side Information, Pre-interpolation, Data Embedding.
\end{abstract}

\section{Introduction}

\label{sec:intro} Image restoration is intended to compensate for the loss occurred during transmission or storage. In many applications, image pixels are sampled in a non-uniform manner or it may frequently happen that some pixels are lost or unavailable. A common missing pattern is block loss. Block loss may occur due to the packet loss in image transmission through networks or it may appear in the distorted biomedical signals, corrupted archived images, and erroneously received block-based codes. In this paper, we propose a method for image block loss restoration.

Most of the natural signals, including images, have a sparse representation in transform domains such as discrete fourier transform, discrete cosine transform, and discrete wavelet transform. This property has found wide applications in different fields including signal compression \cite{Mallat_Wavelet_Tour_08}, classification \cite{Elhamifar_Sparse_Sub_Clust_09}, pattern recognition \cite{Wright_Sparse_Rep_Comp_Vis_10}, and denoising \cite{Fletcher_Denoising_06}.

One straightforward application of the sparsity property is sampling and compression. The sparsity based sampling techniques can be divided into two major groups of Compressive Sampling (CS) and random sampling. CS techniques refer to taking a small number of linear measurements of the signal entries with rates significantly lower than the Shannon-Nyquist rate \cite{Candes_Rob_Uncert_Exact_06,Candes_Stable_Sig_06,Candes_Quant_Rob_06,Donoho_Compress_Sens_06}. The original signal is recovered from its compressive measurements using CS recovery techniques, such as Basis Pursuit algorithms \cite{ChenAtomic_Decomp98}, greedy techniques \cite{MallatAdaptiveTF_Decomp1994,TroppRecoveryPartialOrthogonalPursuit2006,NeedellUnifromUncertRegOrthMatching2009,NeedellRecoveryIncompleteRegularized2010,NeedellIterativeSignal2009} and thresholding methods \cite{BlumensathIterSparsApprox2008,FornasierIterThreshAlg2008,AzghaniSampta2013}. %The BP algorithms are based on convex optimization and offer better recovery at the cost of higher computational complexity \cite{ChenAtomic_Decomp98}. The greedy techniques such as Orthogonal Matching Pursuit (OMP) \cite{MallatAdaptiveTF_Decomp1994,TroppRecoveryPartialOrthogonalPursuit2006}, regularized Orthogonal Matching Pursuit (ROMP) \cite{NeedellUnifromUncertRegOrthMatching2009,NeedellRecoveryIncompleteRegularized2010}, and Compressive Sampling Matching Pursuit (CoSaMP) \cite{NeedellIterativeSignal2009} have relatively lesser computational complexity and lower recovery performance. The iterative thresholding methods lie between these two groups in both recovery performance and computational complexity. Some of the well-known iterative thresholding methods are Iterative Hard Thresholding (IHT) \cite{BlumensathIterSparsApprox2008,FornasierIterThreshAlg2008} and Iterative Method with Adaptive Thresholding for Compressed Sensing (IMATCS) \cite{AzghaniSampta2013}.
In random sampling techniques, a subset of the signal entries is selected randomly. The random sampling recovery techniques such as Iterative Method with Adaptive Thresholding (IMAT) \cite{MarvastiAminiUnifiedAppr2012} and Iterative Method with Adaptive Thresholding and Interpolation (IMATI) \cite{AzghaniSampta2013} exploit the sparsity property to interpolate the non-uniform pixels. Random sampling techniques can be also employed for missing pixels restoration.

Other image properties are also used for missing pixels restoration. In \cite{GuleryuzNonlinearApprox2006}, Guleryuz developed adaptive linear estimators for non-stationary signals which can be used for robust estimation of image missing regions. In \cite{TakedaKernelRegression2007}, Takeda et al. presented an adaptive kernel regression method for image interpolation and proposed an iterative scheme to further improve its performance. The problem of image block loss restoration is also related to the image inpainting \cite{bertalmio2003simultaneous,criminisi2004region,elad2005simultaneous,fadili2009inpainting,xu2010image}. The aim of image inpainting is, however, different from image restoration as inpainting mainly focuses on how to fill in the missing or damaged regions of the image in an undetectable manner.

In this paper, we propose an image block loss restoration method, assuming that the recipient is aware of the sparsity pattern as side information of the transmitted image. The corrupted image is restored by consecutively imposing the spatial and transform constraints on the image until a stopping criterion is satisfied. Before the iterations, an image interpolation algorithm is employed to speed up the convergence. The pre-interpolation and the suggested stopping criterion can be also adopted to other iterative methods for performance improvement.

In some applications, the side information may not be available in the receiver. To deal with such situations, we develop a technique to embed the sparsity pattern into the sparse image. In this technique, we first compress the sparsity pattern by a binary image compression algorithm and then embed its coded version in the least significant bits of the image pixels. This technique ensures error-free transmission of the sparsity pattern, at the cost of only a small perturbation on the transmitted image. We discuss the effect of using inaccurate spatial and transform information resulting from data embedding on the performance of the iterative method. Extensive simulations are performed to evaluate the method and investigate the efficiency of the introduced techniques. Simulation results verify that the proposed method outperforms other algorithms for image block loss restoration.

The rest of this paper is organized as follows. Section \ref{sect:prel} provides the definitions and the essential background for the paper. The proposed method is explained in Section \ref{sect:method}. Section \ref{sect:effect} discusses the effect of the data embedding on the image restoration. The simulation results are provided in Section \ref{sect:simulation} and Section \ref{sect:conc} concludes the paper.

\section{Preliminaries}\label{sect:prel}

In this section, we provide the notations, definitions, and the essential background for the paper.

\subsection{Notations}

Images are represented by $N$-dimensional vectors and the transform is implemented using an $N\times N$ matrix $\Phi$, denoting the two-dimensional Discrete Cosine Transform (DCT). DCT is chosen because it offers superior energy compaction properties for images and has been employed as a standard tool in many signal and image processing applications \cite{RaoYipDCT2007}.

\subsection{Definitions}

\subsubsection*{a- Sparsity Projector}

An image is called {\em sparse} if a large portion of its transform coefficients is zero.
%The binary map of zero and nonzero coefficients is the {\em sparsity pattern} and the {\em sparsity rate} is the rate of zero coefficients.
The following is the algorithm for {\em sparsifying} the image $w$ with $r\%$ {\em sparsity rate}.
\begin{center}
 \begin{algorithm}[!htp]
    \caption{Image Sparsifying} \label{alg:SP}
	\begin{algorithmic}[1]
    \State \textbf{Input:} Image $w$ and sparsity rate $r\%$.
	\State \textbf{Output:} Sparse Image $w_s$.

        \State $W \leftarrow \Phi w$
        \State $Y \leftarrow sort(|W|)$
        \State $Threshold \leftarrow Y_{\lfloor \frac{rN}{100}\rceil}$

        \For{ $i = 1, ..., N$ }
            \If{ $|W_i| < Threshold$ }
                \State $W_i \leftarrow 0$
            \EndIf
        \EndFor

        \State $w_s \leftarrow \Phi^{-1}W$
        \State \Return $w_s$

    \end{algorithmic}
 \end{algorithm}
\end{center}

%\begin{align}
%& \nonumber W \leftarrow \Phi w\\
%& \nonumber Y \leftarrow sort(|W|)\\
%& \nonumber Threshold \leftarrow Y_{\lfloor \frac{rN}{100}\rceil}\\
%& \nonumber \mbox{\textbf{for }} i \mbox{\textbf{ from }} 1 \mbox{\textbf{ to }} N \mbox{\textbf{ do }}\\
%& \nonumber  \hspace{1cm} \mbox{\textbf{if }} \mbox{$|W_i| < Threshold$} \\
%& \nonumber   \hspace{2cm} \mbox{$W_i \leftarrow 0$}\\
%& \nonumber \hspace{1cm} \mbox{\textbf{end if}} \\
%& \nonumber \mbox{\textbf{end for}}\\
%& \nonumber w_s \leftarrow \Phi^{-1}W
%\end{align}
In Algorithm \ref{alg:SP}, $|A|$ denotes the absolute value of the vector $A$, $sort(A)$ refers to the sorted version of the vector $A$ in ascending order, $A_i$ is the $i$-th element of the vector $A$ and $\lfloor a \rceil$ denotes the nearest integer to the scalar $a$.

Let $W=\Phi w_s$. The {\em Sparsity Pattern} (SP) of the sparse image $w_s$ is obtained as follows:

\begin{equation}
SP_i =
\begin{cases}
0  & \mbox{ if $W_i=0$} \\
1  & \mbox{ if $W_i\neq 0$} \\
\end{cases}
, 1 \leq i \leq N.
\end{equation}

The sparsity mask $Z$ is an $N\times N$ sparse binary matrix, in which only some of the diagonal elements are one (corresponding to the nonzero SP elements). The Sparsity Projector ($P_{S}$) takes an image $w$ and projects it onto the set of sparse images with the sparsity mask $Z$ as follows:
\begin{equation}\label{eq:P-S}
P_{S}(w)_Z=\Phi^{-1}Z\Phi w.
\end{equation}
Equation~\ref{eq:P-S} implies that $P_{S}$ takes an image and yields the corresponding sparse image with the pre-determined sparsity pattern. With a fixed sparsity pattern, the sparsity projection is a linear operation. Therefore, we can rewrite~\ref{eq:P-S} as:
\begin{equation}\label{eq:S}
x_{s}=Sx
\end{equation}
where $x$ and $x_{s}$ are the original and sparse images, respectively, and $S$ is an $N\times N$ matrix, representing the sparsity projector.

\subsubsection*{b- Data Projector}
In the missing pixel model, a pixel is either delivered flawlessly or corrupted entirely, respectively named as known and missing pixels. Therefore, we can write:
\begin{equation}
y=Mx
\end{equation}
where $x$ and $y$ are the original and corrupted images, respectively, and $M$ is the sampling mask. The sampling mask is an $N\times N$ sparse binary matrix, in which only some of the diagonal elements are one (corresponding to the position of known pixels).

The Data Projector ($P_{D}$) takes an image $w$ and, using the sampling mask $M$, projects it onto the image $y$ as follows:

\begin{equation}\label{eq:P-D1}
P_{D}(w)_{M,y}=\hat{M}w+My
\end{equation}
where $\hat{M}=(I-M)$ and $I$ is the identity matrix.
The effect of the data projector on the image $w$ is:
\begin{equation}\label{eq:P-D2}
\mbox{Effect of }P_{D}=P_{D}(w)_{M,y}-w\\
=(I-M)w+My-w=M(y-w).
\end{equation}
Equations~\ref{eq:P-D1} and~\ref{eq:P-D2} indicate that $P_D$ resets the uncorrupted pixels to their original value and leaves other pixels unchanged.
%substitutes pixels with their original values if they are in the corresponding positions of the known pixels and leaves other pixels unchanged.

\subsection{Binary Erasure Channel}

\begin{figure}[h]
\centering
\includegraphics[width=0.35\textwidth]{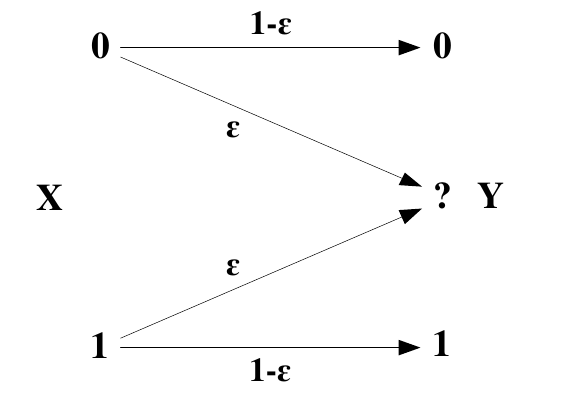}
\caption{The Binary Erasure Channel with error probability $\epsilon$.}\label{fig:BEC}
\end{figure}

Binary Erasure Channel (BEC) is a common communication channel model. In this model, the recipient receives either the transmitted bit or a message that the bit is erased. Therefore, we can model the missing pixels as the BEC. Figure \ref{fig:BEC} illustrates a BEC with error probability $\epsilon$.

\subsection{Low-Density Parity-Check Coding}\label{sect:LDPC}

Low-Density Parity-Check (LDPC) code is a linear error correcting code used for protecting messages during transmission over noisy channels \cite{MacKay2003,MoonWiley2005}, and is constructed using a sparse bipartite graph \cite{ShokLDPC2003}. LDPC codes are capacity-approaching codes, which means that practical construction methods exist that allow the noise threshold to be set very close (or even arbitrarily close for the BEC) to the theoretical maximum (the Shannon limit) for a symmetric memory-less channel \cite{luby2001efficient}. The noise threshold defines an upper bound for the channel noise, where the loss probability can be arbitrarily small. 
\section{THE PROPOSED METHOD}\label{sect:method}

\begin{figure*}[h]
\centering
\begin{subfigure}[b]{0.81\linewidth}
    \includegraphics[width=\linewidth]{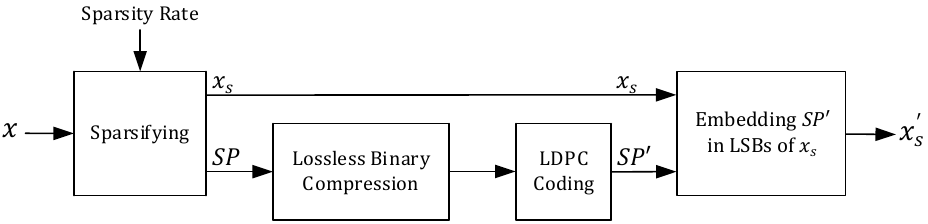}
		  \caption{}
\end{subfigure}
\begin{subfigure}[b]{0.31\linewidth}
    \includegraphics[width=\linewidth]{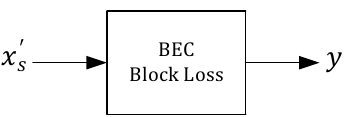}
		  \caption{}
\end{subfigure}
\begin{subfigure}[b]{0.96\linewidth}
    \includegraphics[width=\linewidth]{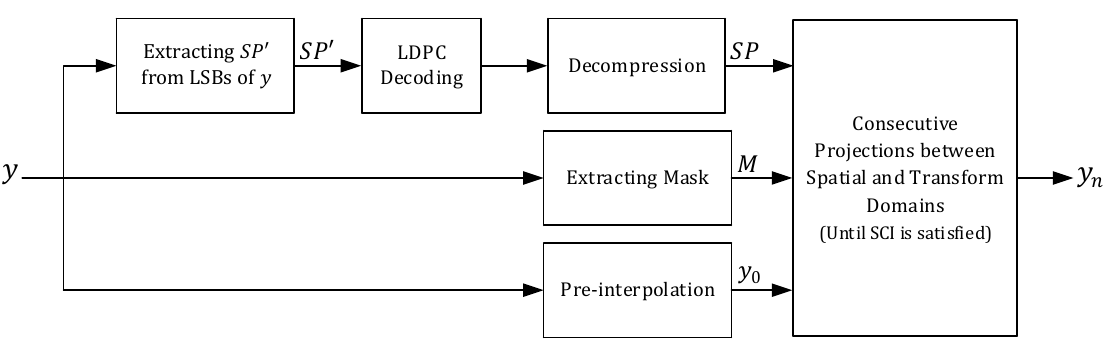}
  \caption{}
\end{subfigure}
\caption{The block diagram of the image restoration method. (a) Transmitter, where $x$ and $x_s$ are the original and sparse images, respectively, $SP$ is the sparsity pattern, $SP^{'}$ is the LDPC coded version of the compressed $SP$ and $x^{'}_s$ is the sparse image with $SP^{'}$ embedded in the LSBs, (b) Channel, where $y$ is the corrupted image, and (c) Receiver, where $y_0$ is the interpolated version of $y$, $y_n$ is the restored image and sparsity and data projectors are obtained from the sparsity pattern $SP$ and the sampling mask $M$, respectively.}\label{fig:method}
\end{figure*}

In this section, we propose a method for image block loss restoration, assuming that the sparsity pattern of the transmitted image is available as side information. The block diagram of the method is depicted in Figure~\ref{fig:method}.

\subsection{The Basic Iterative Method}

For image restoration, we use the concept of error correction in channel coding theory. It is known that, to achieve error correction, one can add a kind of redundancy to the message which is exploited in the decoder to recover the data. In our case, the sparsity projector provides linear equations which can be used for error correction.

Suppose that the sparse image is transmitted via a channel that causes block loss and the knowledge of the sparsity pattern is available at the receiver. The recipient restores the corrupted image using the spatial and transform domains constraints, which are the position and value of known pixels and the position of transform zeros, respectively. If the original image is sparse enough, the consecutive projections between these two domains yield the transmitted sparse image.

Let $x$ and $x_s$ be the original and transmitted sparse images and $y$ be the corrupted image. Let $Z$ be sparsity mask of $x_s$ and $M$ be the sampling mask induced by the channel. The iterative method is formulated as follows:

\begin{equation}\label{eq:iter1}
y_{k}=P_{D}(P_{S}(y_{k-1})_Z)_{M,y}, k \geq 1,
\end{equation}
where $y_k$ is the result of the $k$-th iteration and $y_{0}=y$.
According to~\ref{eq:S} and~\ref{eq:P-D1}, we can rewrite~\ref{eq:iter1} as:
\begin{equation}\label{eq:iter2}
y_{k}=(\hat{M}S)y_{k-1}+My_{0}, k \geq 1.
\end{equation}
The recursive relation of~\ref{eq:iter2} yields the following solution:
\begin{equation}
\begin{split}
  y_{k} & =(\hat{M}S)^{k}y_{0}+\Sigma_{i=0}^{k-1}{(\hat{M}S)^i(My_{0})} \\
  &=(\hat{M}S)^{k}y_{0}+(I-(\hat{M}S)^{k})(I-\hat{M}S)^{-1}My_{0}, k \geq 1.
\end{split}
\end{equation}
Since $Sx_{s}=x_{s}$ and $My_{0}=Mx_{s}$, we have:
\begin{equation}\label{eq:My0}
My_{0}=(I-\hat{M})x_{s}=x_{s}-\hat{M}Sx_{s}=(I-\hat{M}S)x_{s}.
\end{equation}
Therefore
\begin{equation}\label{eq:iter3}
\begin{split}
  y_{k}&=(\hat{M}S)^{k}y_{0}+(I-(\hat{M}S)^{k})(I-\hat{M}S)^{-1}(I-\hat{M}S)x_{s}\\
 & =(\hat{M}S)^{k}y_{0}+(I-(\hat{M}S)^{k})x_{s}\\
 & =x_{s}+(\hat{M}S)^{k}(y_{0}-x_{s}).
\end{split}
\end{equation}

From~\ref{eq:iter3}, it is inferred that the deviation of the result of the $k$-th iteration from the transmitted sparse image is $(\hat{M}S)^{k}(y_{0}-x_{s})$. Since the term $(y_0-x_s)$ is constant, the convergence rate is related to the matrix $(\hat{M}S)$. It implies that the performance of the iterative method depends on the missing probability and the sparsity rate of the transmitted image; This means that for an image with more sparsity, the recipient can restore larger losses. However, it should be noted that although a higher sparsity rate allows faster convergence to the sparse image, it tends to blur the original image. Since our aim is the reliable transmission of the original image, we should decide on the trade-off between the amount of image blurring and the capability of image restoration according to the channel parameters.

\subsection{Improving the Performance of the Iterative Method}

Two features, including a pre-interpolation and a criterion for stopping the iterations, are introduced to improve the performance of the proposed method. The features are described in the following.

\subsubsection*{a- Pre-Interpolation}

Due to the loss of a portion of the received image in the spatial domain, the coefficients of the transform domain, including the DC coefficient, does not reflect their true values. Thus, for each pixel, it takes several iterations to reach a proper value, particularly when a larger number of pixels are affect by the block loss. Interpolation provides a better starting point for the iterations, which significantly increases the convergence speed. The pre-interpolation also helps the method to converge when the sparsity rate is not enough to restore the block losses. The impact of pre-interpolation can be also observed in~\ref{eq:iter3}; the restored image reaches $x_{s}$ in fewer iterations if $y_{0}$ provides a better approximation of $x_{s}$.

For the pre-interpolation, we need a fast method to efficiently estimate the original values of the missing pixels. For this purpose, we employ the Adaptive Iterative Mean filter (AIM) introduced in \cite{Hosseini2013}; AIM iteratively restores missing pixels, where the number of iterations is different for each missing pixel, depending on the Euclidean distance from the nearest uncorrupted pixel.

\subsubsection*{b- Stopping Criterion for Iterations}

We suggest an efficient approach to stop the iterations. The approach is to identify when more iterations bring no improvement. Suppose that $y_{k}^{S}=P_{S}(y_{k-1})_Z$. According to~\ref{eq:P-D2}, the effect of the data projector at iteration $k$ is as follows:

\begin{equation}
\Delta y{}_{k}=M(y_{0}-y_{k}^{S}).
\end{equation}
The known pixels of $y_{0}$ have integer values. Thus, if the absolute values of the elements of $\Delta y_{k}$ lie in $[0\,0.5)$, the data projector is equivalent to rounding the pixel values. As a result, it becomes independent of $y_{0}$ and proceeding the iterations does not make further improvement. In this case, in the corresponding position of the known pixels, the absolute value of $\Delta y_{k}$ has a uniform distribution with the mean value of $0.25$. Thus, we define the Stopping Criterion for Iterations as follows.

\noindent \textbf{Stopping Criterion for Iterations (SCI): }\emph{Stop the iterations if the data projector causes the average change of lower than $0.25$ in the corresponding position of the known pixels.}

\noindent Generally, as the sparsity rate increases, the SCI is satisfied in fewer iterations.

\subsection{Embedding the Sparsity Pattern in the Sparse Image}\label{sect:embed}

In many applications, the availability of the sparsity pattern at the receiver may not be a realistic assumption. In this following, we develop a technique for transmitting the sparsity pattern along with the sparse image.

\subsubsection*{a- Compressing the Sparsity Pattern}\label{sect:comp}

Due to the smoothness of the natural images, higher frequencies are much more probable to be zero. As a result, the sparsity pattern is not a random binary matrix and, thus, is compressible with the aid of efficient binary image compression algorithms. Assuming a compression rate of $0<c<1$, the sparsity pattern of an $N$-dimensional image can be represented by $cN$ bits.% and consequently a number of $\frac{cN}{1-\epsilon}$ bits should be embedded into the sparse image. This means that, on average, a number of $\frac{c}{1-\epsilon}$ LSBs of each pixel will be devoted to the sparsity pattern, which reduces the distortion of the transmitted image caused by data embedding.

\subsubsection*{b- Using LDPC Coding to Protect the Compressed Sparsity Pattern}\label{sect:code}

It is known that the channel capacity of the BEC, with error probability $\epsilon$, is $1-\epsilon$ \cite{ShokLDPC2003}. Therefore, to transmit the compressed sparsity pattern of an $N$-dimensional image, a number of $\frac{cN}{1-\epsilon}$ bits should be embedded into the sparse image. To minimize the distortion introduced to the sparse image, we embed these extra bits in the pixel LSBs. This means that, on average, a number of $\frac{c}{1-\epsilon}$ LSBs of each pixel will be devoted to the sparsity pattern. To map the $cN$ bits of the sparsity pattern to $\frac{N}{1-\epsilon}$ bits, we use LDPC coding because, as mentioned in subsection \ref{sect:LDPC}, theses codes are known to be capacity-approaching for the BEC. The resultant is interleaved to overcome the burst error caused by block losses.

Since the data embedding affects the value of the sparse image pixels, we have to modify the data projector in the restoration algorithm. Previously, the data projector forced the known pixels to hold their actual value in each iteration; now it replaces just those bits that are not affected by embedding the sparsity pattern. In other words, the data projector replaces all bits except the LSBs that were devoted to the sparsity pattern. However, although the SCI is closely related to the data projector, its definition does not change. That is we stop the iterations if the modified data projector causes the average change of lower than $0.25$ in the corresponding position of the known pixels.

\section{THE EFFECT OF DATA EMBEDDING ON THE RESTORATION STAGE}\label{sect:effect}

Embedding the sparsity pattern in the sparse image degrades the quality of the transmitted sparse image, and, as a result, the spatial information, used in iterations, would be inaccurate. Also, there might be some errors in the decoded sparsity pattern. These effects can potentially harm the performance of the iterative method. In this section, we discuss these unwanted consequences of embedding the sparsity pattern in the sparse image.

\subsection{The effect of Inaccurate Spatial Information}

Assume that a number of $k$ LSBs of the image pixels are devoted to the sparsity pattern. The values of both the $k$ original and modified LSBs are discrete random variables with a uniform distribution function, say $d$, and elements in $[0,2^{k})$; thus, the change in the $k$ image LSBs is a random variable $D$ with the distribution function $d*(-d)$ where the symbol $*$ is the convolution operation. This random variable has the mean value of zero and the variance of $\sigma^{2}=\frac{2^{k}(2^{k-1}+1)}{6}$. The modified sparse image is ${x_{s}}^{D}=x_{s}+D$ and we can rewrite~\ref{eq:My0} as follows:
\begin{equation}
My_{0}=Mx_{s}^{D}=M(x_{s}+D)=(I-\hat{M}S)x_{s}+MD.
\end{equation}
Therefore
\begin{equation}
y_{k}=x_{s}+(\hat{M}S)^{k}(y_{0}-x_{s})+(I-(\hat{M}S)^{k})(I-\hat{M}S)^{-1}MD.
\end{equation}
Thus, the error caused by altering the image LSBs is equal to $(I-(\hat{M}S)^{k})(I-\hat{M}S)^{-1}MD$. Assuming that $\|\hat{M}S\|<1$, this error is approximately equal to $(I-\hat{M}S)^{-1}MD$. It implies that the method  converges to an image which is contaminated by a zero mean noise with a variance of $O(\sigma^{2})$. This error is the trade-off for not to consider the sparsity pattern as shared information between the transmitter and the receiver. Although this error generally degrades the quality of the restored image, simulation results confirm that the degradation is quite negligible.

\subsection{The effect of Inaccurate Sparsity Pattern}\label{sect:effectSP}

For inaccurate sparsity pattern, we consider the cases of false negative (undetected zero coefficients) and false positive (coefficients identified as zero in error). Assume that $S^{N}$ and $S^{P}$ are matrix representations of the sparsity projector, corresponding to sparsity patterns with false negative and false positive, respectively. We analyze these cases in the following.

\subsubsection*{a- False Negative}

In this case, we have $S^{N}x_{s}=x_{s}$. Therefore, Equations~\ref{eq:My0} and~\ref{eq:iter3} hold and the method converges, although at a slower rate.

\subsubsection*{b- False Positive}

In this case, we have $S^{P}x_{s}\neq x_{s}$. Assume that $S=S^{P}+S^{\delta}$ where $S^{\delta}$ is the matrix representation of the sparsity projector, corresponding to the transform coefficients which are wrongly detected as zero. We can rewrite~\ref{eq:My0} as follows:
\begin{equation}
\begin{split}
 My_{0} &=(I-\hat{M})x_{s}=x_{s}-\hat{M}(S^{P}+S^{\delta})x_{s}\\
 & =(I-\hat{M}S^{P})x_{s}-\hat{M}S^{\delta}x_{s}.
\end{split}
\end{equation}
Therefore, Equation~\ref{eq:iter3} will be modified as:
\begin{equation}
\begin{split}
  y_{k}&=x_{s}+(\hat{M}S^{P})^{k}(y_{0}-x_{s})-(I-(\hat{M}S)^{k})(I-\hat{M}S^{P})^{-1}\hat{M}S^{\delta}x_{s}.
\end{split}
\end{equation}
Thus, the error caused by the false positive is equal to $(I-(\hat{M}S^{P})^{k})(I-\hat{M}S^{P})^{-1}\hat{M}S^{\delta}x_{s}$. Assuming that $\|\hat{M}S^{P}\|<1$, this error is approximately equal to  $(I-\hat{M}S^{P})^{-1}\hat{M}S^{\delta}x_{s}$. This term does not fade with proceeding the iterations and introduces an error floor for the restored image. This effect emphasizes that the method is very sensitive to the wrong-detection of non-zero coefficients as zero. Therefore, to attain a reliable restoration, the rate of LDPC coding should be chosen carefully to ensure error-free transmission of the sparsity pattern.

\section{SIMULATION RESULTS}\label{sect:simulation}

Extensive simulations are performed to investigate the performance of the proposed method and evaluate the efficiency of the introduced techniques. Simulations are carried out on several 8-bit gray-scale images. The results are illustrated for three $512 \times 512$ test images \emph{Lena}, \emph{Boat} and \emph{Baboon}. We consider missing patterns as block losses with block sizes of $1$, $2$, $4$, $8$ and $16$. The block loss with block size of one is also called the random loss. For better comparison, the missing probability has been fixed at $50\%$. The PSNR value of the restored image with respect to the sparse and original images are denoted by PSNR-RS and PSNR-RO, respectively. In the following, we provide the simulation results.

\subsection{The Effect of Pre-interpolation}

Using the pre-interpolation, the method converges on average three times faster. It also slightly improves the PSNR-RO and PSNR-RS values. More importantly, with pre-interpolation the method is more likely to converge especially when the sparsity rate is not enough to restore the block losses. Table~\ref{table:T-Preint} investigates the effect of pre-interpolation on block loss restoration for the sparse images with different sparsity rates.
While without the pre-interpolation there are strict requirements for the minimum sparsity rate to satisfy the SCI, the method always converges using the pre-interpolation. For the 95\% sparsity rate, in which the method converges for all cases, the required number of iterations is mentioned in the table to indicate the effect of pre-interpolation on the convergence speed.

\begin{table}%[h]
\centering
\caption{PSNR-RS (dB) values for restoring the sparse images, with different sparsity rates, that are corrupted by $50\%$ block loss with block sizes of $8$. The required number of iterations to satisfy SCI is specified for the case of the 95\% sparsity rate. (In cases that the method does not converge, the results are shown for $1000$ iterations.)}\label{table:T-Preint}
\begin{tabular}{|c|c|c|c|c|c|c|c|} \hline
                                                      & \textbf{65\%} & \textbf{70\%} & \textbf{75\%} & \textbf{80\%} & \textbf{85\%} & \textbf{90\%} & \textbf{95\%}  \\ \hline
\textbf{\emph{Lena} without pre-interpolation}        & 11.4  & 11.9  & 12.4  & 13.9  & 15.5  & 20.2  & 42.7 (953)  \\ %\hline
\textbf{\emph{Lena} with pre-interpolation}           & 28.1  & 28.5  & 29.0  & 30.1  & 31.9  & 35.9  & 44.9 (234)  \\ \hline
\textbf{\emph{Boat} without pre-interpolation}        & 10.3  & 10.9  & 11.5  & 12.6  & 14.8  & 20.3  & 44.2 (827)  \\ %\hline
\textbf{\emph{Boat} with pre-interpolation}           & 26.1  & 26.6  & 27.4  & 28.8  & 31.5  & 36.4  & 45.9 (215)   \\ \hline
\textbf{\emph{Baboon} without pre-interpolation}      & 12.8  & 13.9  & 15.9  & 20.9  & 32.0  & 43.0  & 50.6 (202)  \\ %\hline
\textbf{\emph{Baboon} with pre-interpolation}         & 26.8  & 28.6  & 31.3  & 34.8  & 39.3  & 44.8  & 51.6 (80)   \\ \hline
\end{tabular}
\end{table}

\subsection{The Effect of Data Embedding}

As stated in subsection~\ref{sect:comp}, the sparsity pattern can be compressed using efficient binary image compression algorithms. Using the algorithm introduced in \cite{Kafashan2010}, We can achieve the average compression rate of $c=\frac{1}{2}$. This means that, in the case of the $50\%$ missing probability, a single LSB can bear the LDPC coded version of the compressed sparsity pattern. This will reduce the distortion of the transmitted image that is caused by the data embedding. In the following we investigate the effect of the data embedding on the restoration stage through simulation.

\subsubsection*{a- The Effect of Inaccurate Spatial Information}

Table~\ref{table:T-inacc} shows the effect of using inaccurate spatial information on the restoration stage. It can be seen that, by altering the image LSBs, the restoration quality only slightly degrades; therefore, the method is quite stable with non-exact spatial information.

\begin{table}%[h]
\centering
\caption{PSNR-RO (dB) values for restoring the sparse images, with different sparsity rates, that are corrupted by random loss.}\label{table:T-inacc}
\begin{tabular}{|c|c|c|c|c|c|c|c|} \hline
                                                   & \textbf{65\%}  & \textbf{70\%}  & \textbf{75\%}  & \textbf{80\%}  & \textbf{85\%}  & \textbf{90\%}  & \textbf{95\%}   \\ \hline
\textbf{\emph{Lena} without data embedding}        & 41.1	& 40.3	& 39.2	& 37.9	& 36.4	& 34.4	& 31.5   \\ %\hline
\textbf{\emph{Lena} with data embedding}           & 39.7	& 39.3	& 38.5	& 37.5	& 36.1	& 34.3	& 31.5   \\ \hline
\textbf{\emph{Boat} without data embedding}        & 40.9	& 39.9	& 38.5	& 36.8	& 34.7	& 32.3	& 29.2   \\ %\hline
\textbf{\emph{Boat} with data embedding}           & 39.2	& 38.9	& 37.9	& 36.4	& 34.5	& 32.2	& 29.2   \\ \hline
\textbf{\emph{Baboon} without data embedding}      & 31.3	& 30.1	& 28.8	& 27.5	& 26.1	& 24.6	& 22.8   \\ %\hline
\textbf{\emph{Baboon} with data embedding}         & 31.0	& 29.9	& 28.7	& 27.5	& 26.1	& 24.6	& 22.8   \\ \hline
\end{tabular}
\end{table}

\subsubsection*{b- The Effect of Inaccurate Sparsity Pattern}

Table~\ref{table:T1} investigates the effect of using non-exact sparsity pattern on the restoration stage. Let the modified sparsity pattern be the sparsity pattern that is used in iterations. It can be seen that if the modified sparsity pattern is a subset of the original sparsity pattern the image will be restored, although in more iterations; thus, false negative is tolerable. However, if the original sparsity pattern is a subset of the modified sparsity pattern, the method will not converge. Table~\ref{table:T1} clarifies the sensitivity of the method to the wrongly detected zero coefficients. Therefore, as stated in subsection~\ref{sect:effectSP}, it is necessary to choose a sufficient coding rate to avoid the false positive case in the sparsity pattern during the restoration.

\begin{table}[h]
\centering
\caption{PSNR (dB) values of the restored image \emph{Lena} corrupted by random loss, using non-exact sparsity patterns.}\label{table:T1}
\begin{tabular}{|c|c|c|c|c|}
\hline
\textbf{original sparsity rate}                                                      & \multicolumn{2}{c|}{75\%} & \multicolumn{2}{c|}{76\%} \\ \hline
\textbf{\begin{tabular}[c]{@{}c@{}}sparsity rate \\ used in iterations\end{tabular}} & 75\%   & 76\%             & 75\%        & 76\%        \\ \hline
\textbf{PSNR-RO}                                                                     & 39.2   & No Convergence   & 38.9        & 39.0        \\ \hline
\end{tabular}
\end{table}

%\begin{figure}[h]
%\centering
%\includegraphics[width=0.75\textwidth]{Figures/Fig1.pdf}
%\caption{The effect of pre-interpolation on block loss restoration. The missing pattern is $50\%$ block loss with block sizes of $8$. The required number of iterations to satisfy SCI is specified for the case of the $95\%$ sparsity rate. In cases where the method does not converge, the results are shown for $1000$ iterations.}\label{fig:preint}
%\end{figure}

%\begin{figure}[h]
%\centering
%\includegraphics[width=0.75\textwidth]{Figures/Fig2.pdf}
%\caption{The effect of using inaccurate spatial information on the restoration stage. The results are shown for the three sparse images that are corrupted by random loss.}\label{fig:inaccurate}
%\end{figure}

\subsection{The Method with All Features}

In this subsection, we evaluate the performance of the method that includes all features and compare the results with other methods. Comparisons are done in terms of the PSNR-RO value and the visual quality.

\begin{figure*}[hp]
\centering
\begin{subfigure}[b]{0.55\linewidth}
    \includegraphics[width=\linewidth]{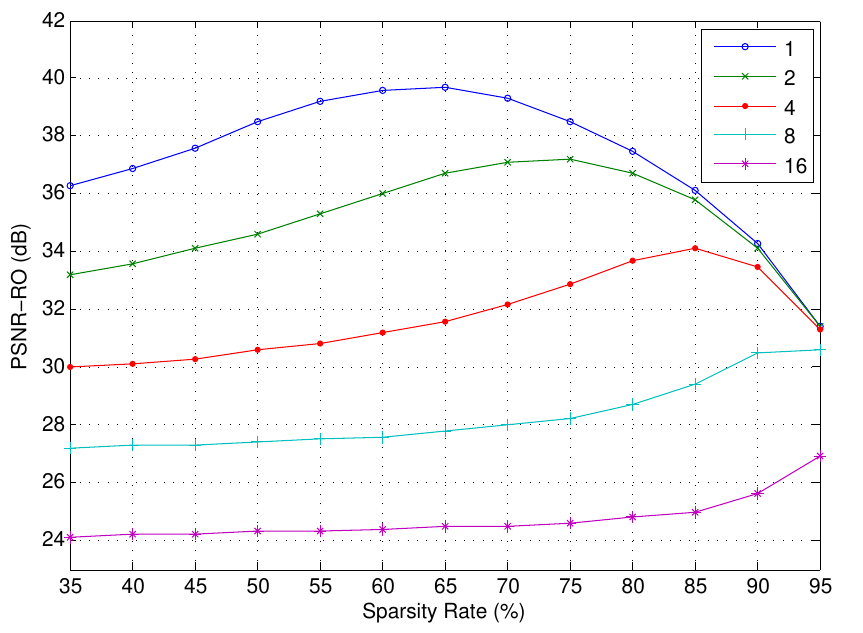}
		  \caption{}
\end{subfigure}
\begin{subfigure}[b]{0.55\linewidth}
    \includegraphics[width=\linewidth]{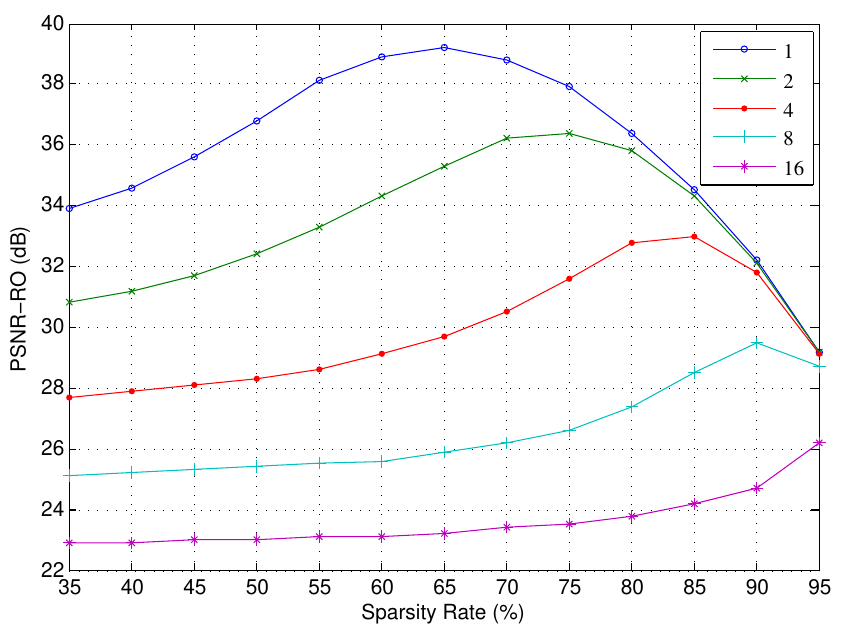}
		  \caption{}
\end{subfigure}
\begin{subfigure}[b]{0.55\linewidth}
    \includegraphics[width=\linewidth]{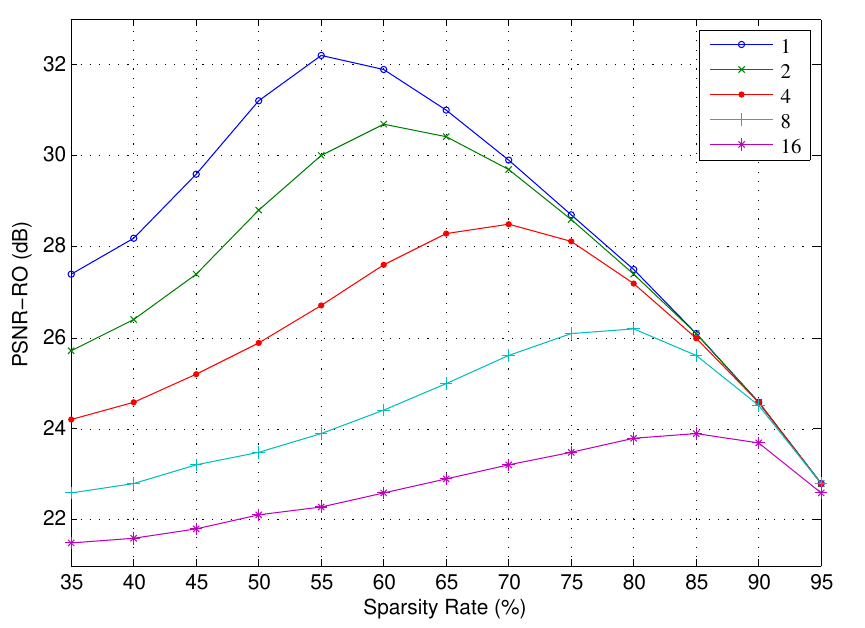}
  \caption{}
\end{subfigure}
\caption{The results for restoring the sparse images, with different sparsity rates, that are corrupted by various block losses. (a) Image \emph{Lena}, (b) Image \emph{Boat}, and (c) Image \emph{Baboon}. The numbers indicate the size of missing blocks.}\label{fig:spar}
\end{figure*}

Figure~\ref{fig:spar} provides the results of the proposed method for restoring the three sparse images, with different sparsity rates, that are corrupted by block loss with various block sizes. It can be seen that, for each image and missing pattern, there is an optimum sparsity rate which yields the best outcome. Generally, the optimum sparsity rate increases as the block size increases. Besides, the results illustrate that, as expected, larger losses are more difficult to restore.

\begin{figure*}[hp]
\centering
\begin{subfigure}[b]{0.55\linewidth}
    \includegraphics[width=\linewidth]{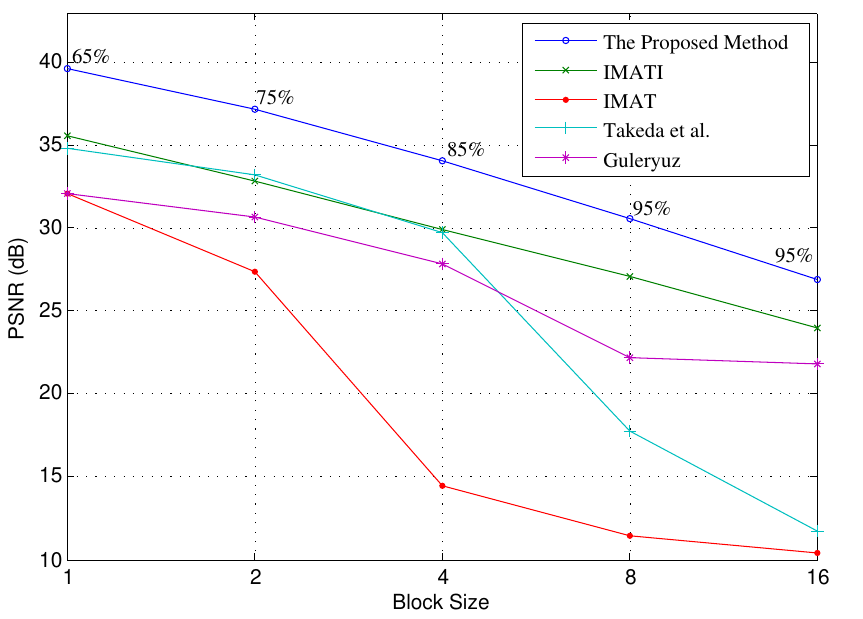}
		  \caption{}
\end{subfigure}
\begin{subfigure}[b]{0.55\linewidth}
    \includegraphics[width=\linewidth]{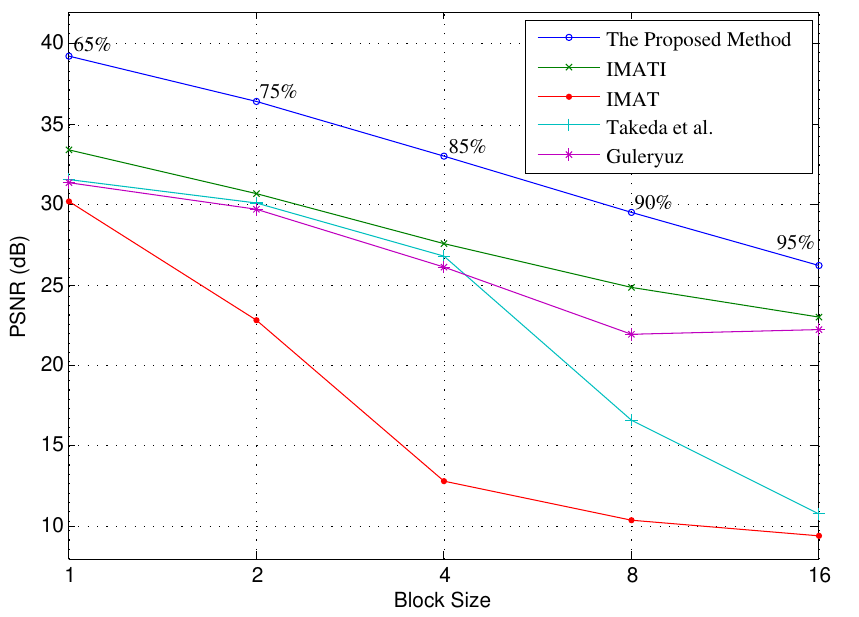}
		  \caption{}
\end{subfigure}
\begin{subfigure}[b]{0.55\linewidth}
    \includegraphics[width=\linewidth]{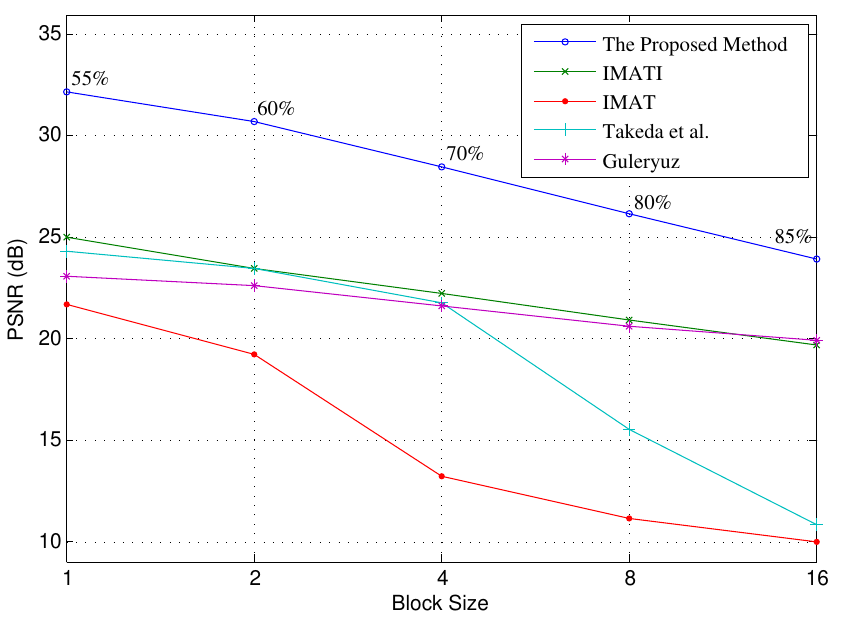}
  \caption{}
\end{subfigure}
\caption{Comparing different methods for restoring the images that are corrupted by various block losses. (a) Image \emph{Lena}, (b) Image \emph{Boat}, and (c) Image \emph{Baboon}. The numbers indicate the sparsity rate of the transmitted sparse image for the proposed method.}\label{fig:block}
\end{figure*}

\begin{figure*}[hp]
\centering
\begin{subfigure}[b]{0.35\linewidth}
    \includegraphics[width=\linewidth]{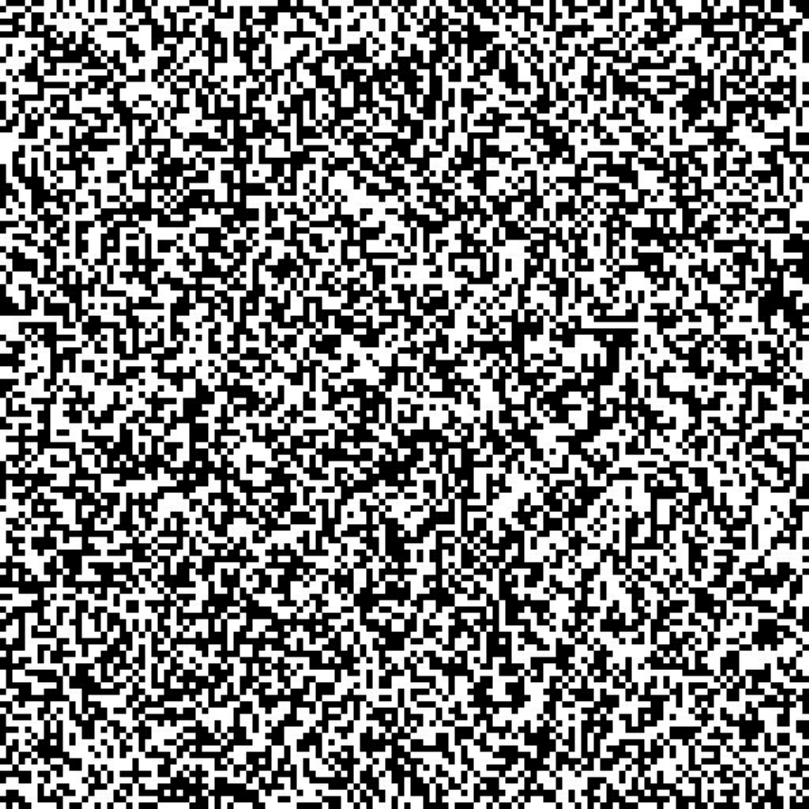}
		  \caption{}
\end{subfigure}
\begin{subfigure}[b]{0.35\linewidth}
    \includegraphics[width=\linewidth]{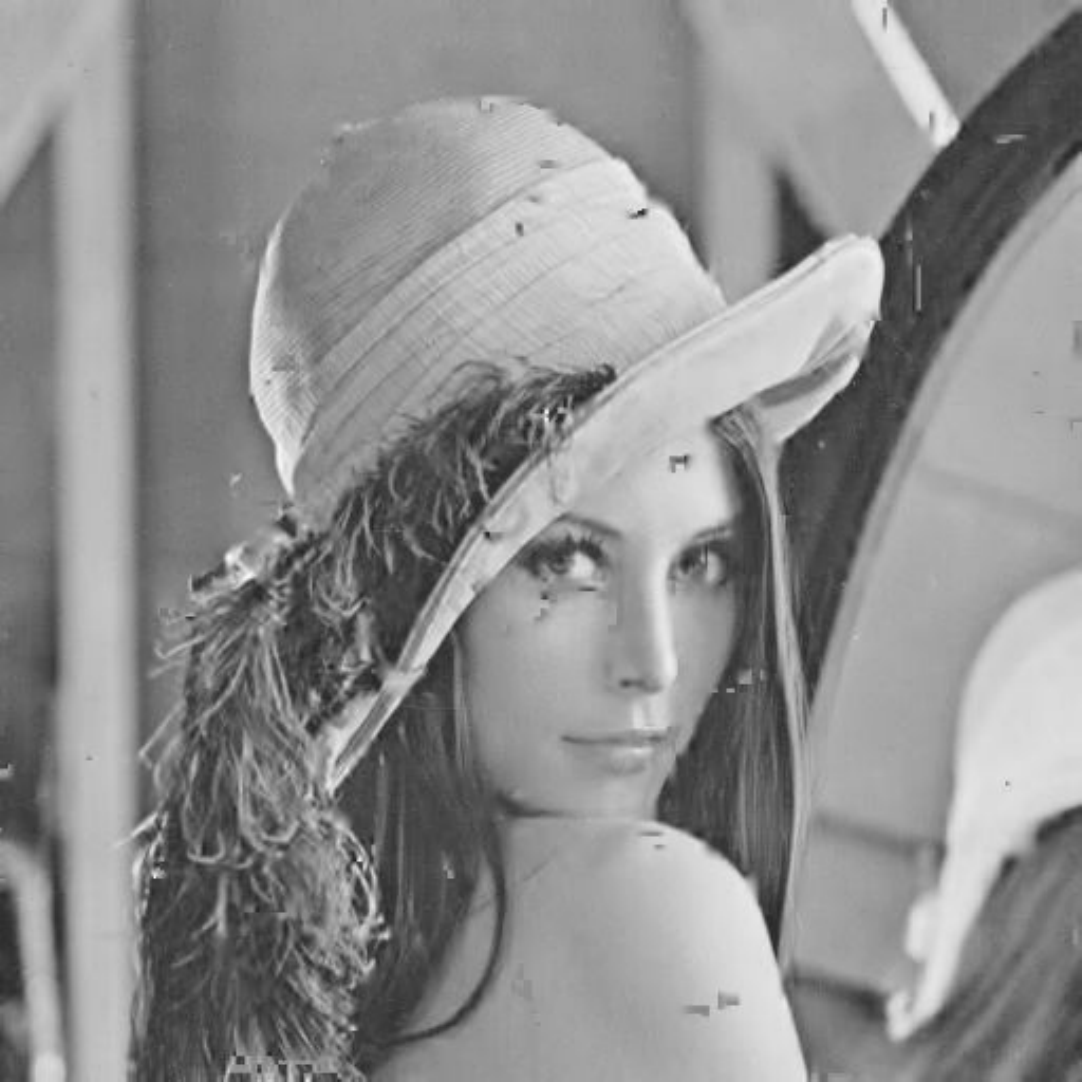}
		  \caption{}
\end{subfigure}
\begin{subfigure}[b]{0.35\linewidth}
    \includegraphics[width=\linewidth]{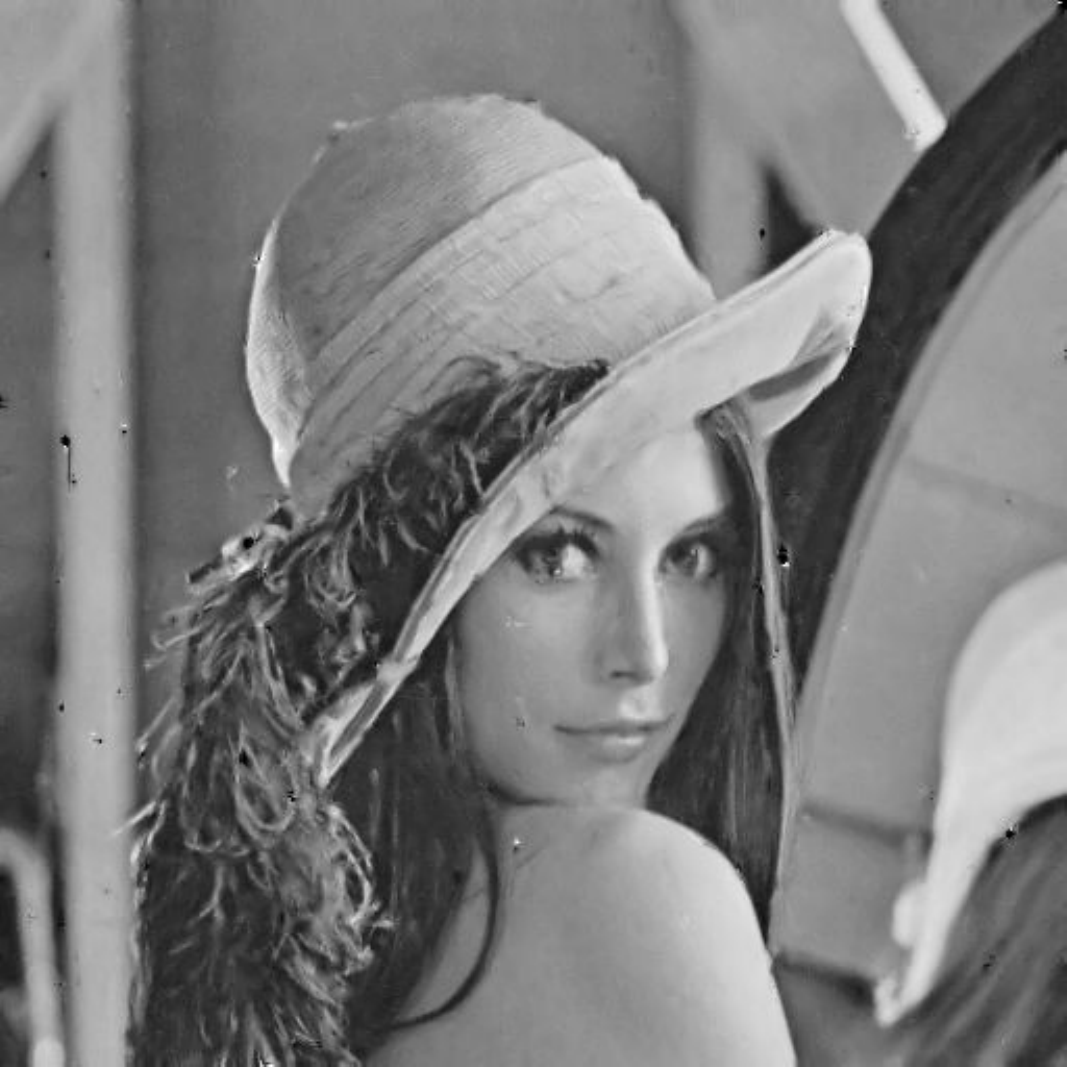}
  \caption{}
\end{subfigure}
\begin{subfigure}[b]{0.35\linewidth}
    \includegraphics[width=\linewidth]{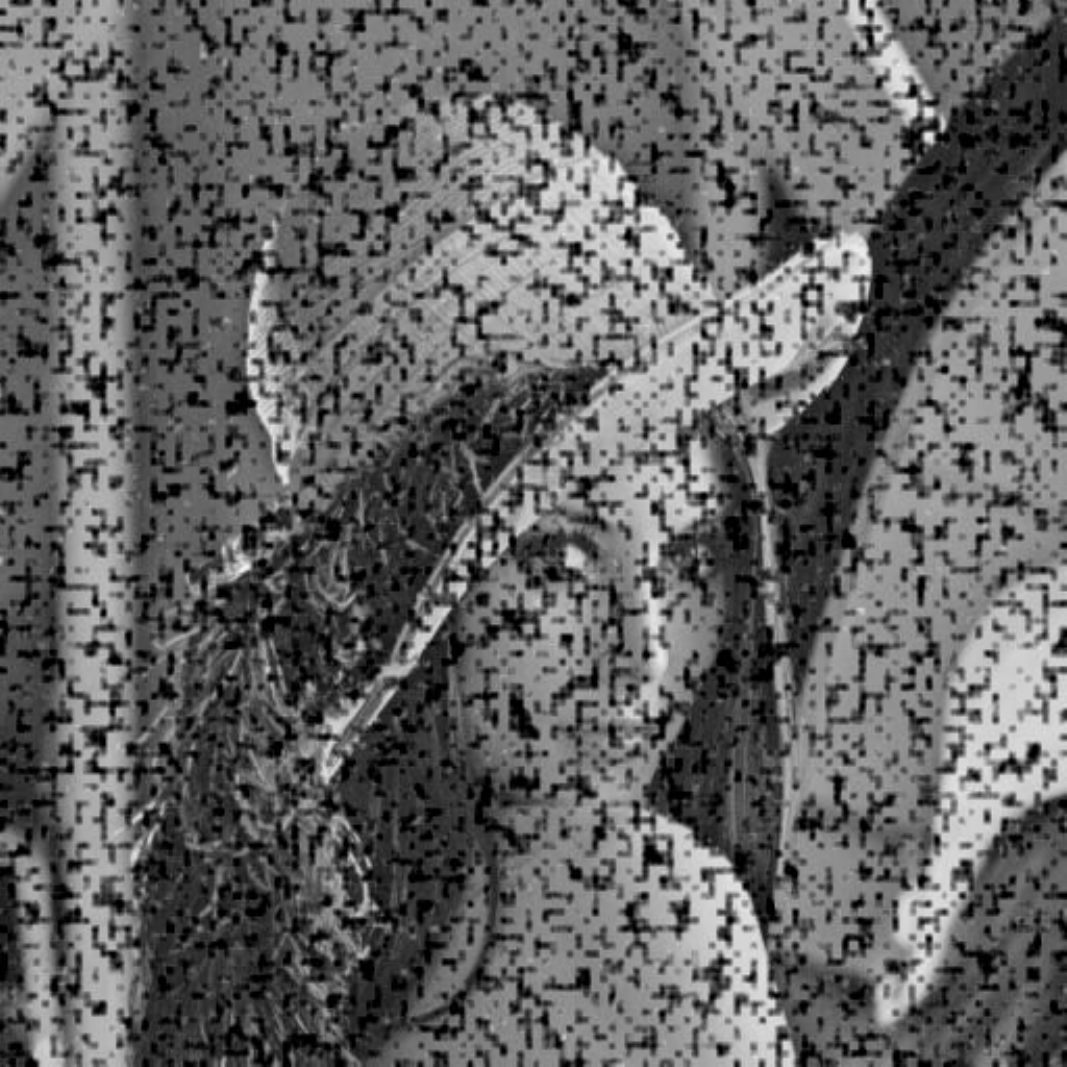}
  \caption{}
\end{subfigure}
\begin{subfigure}[b]{0.35\linewidth}
    \includegraphics[width=\linewidth]{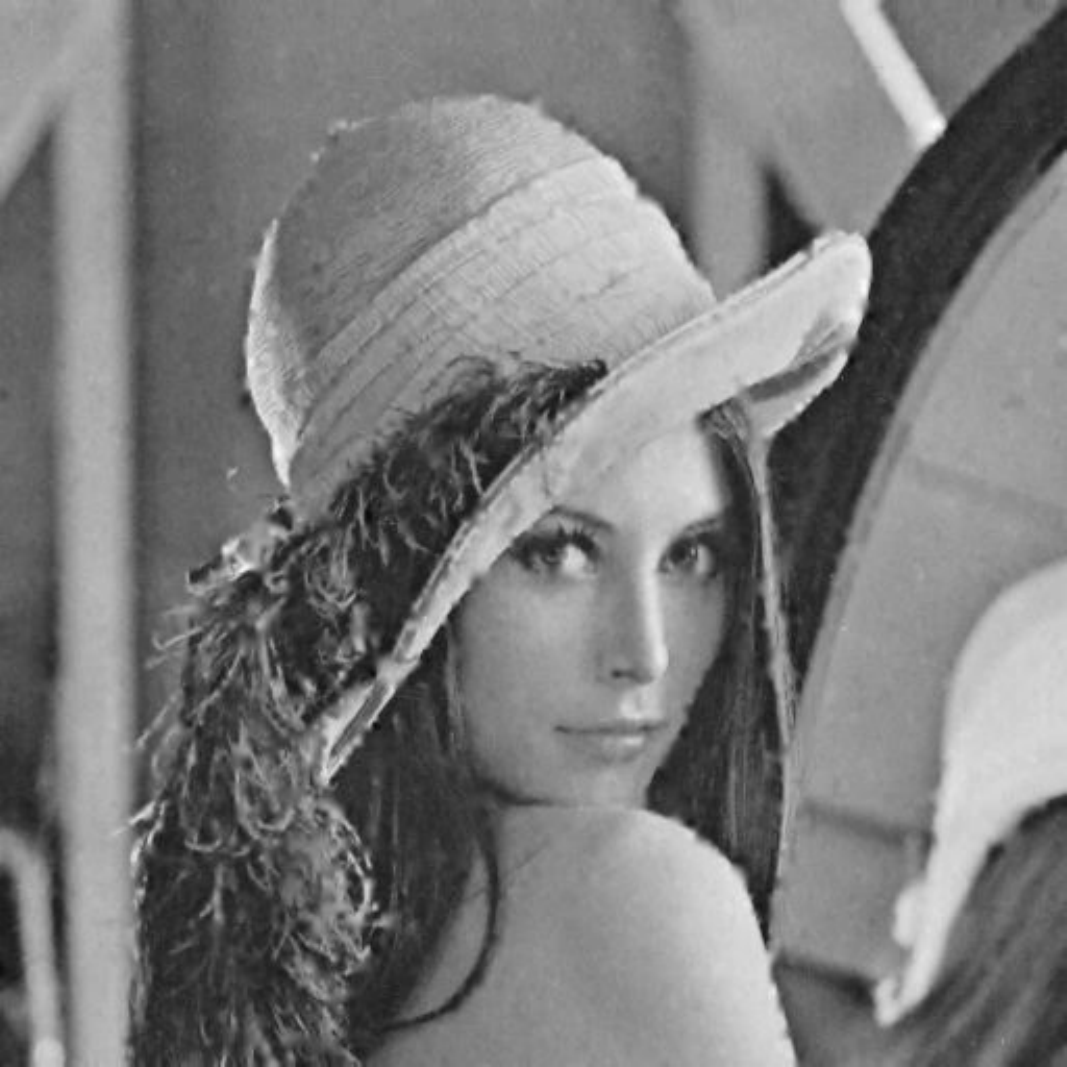}
  \caption{}
\end{subfigure}
\begin{subfigure}[b]{0.35\linewidth}
    \includegraphics[width=\linewidth]{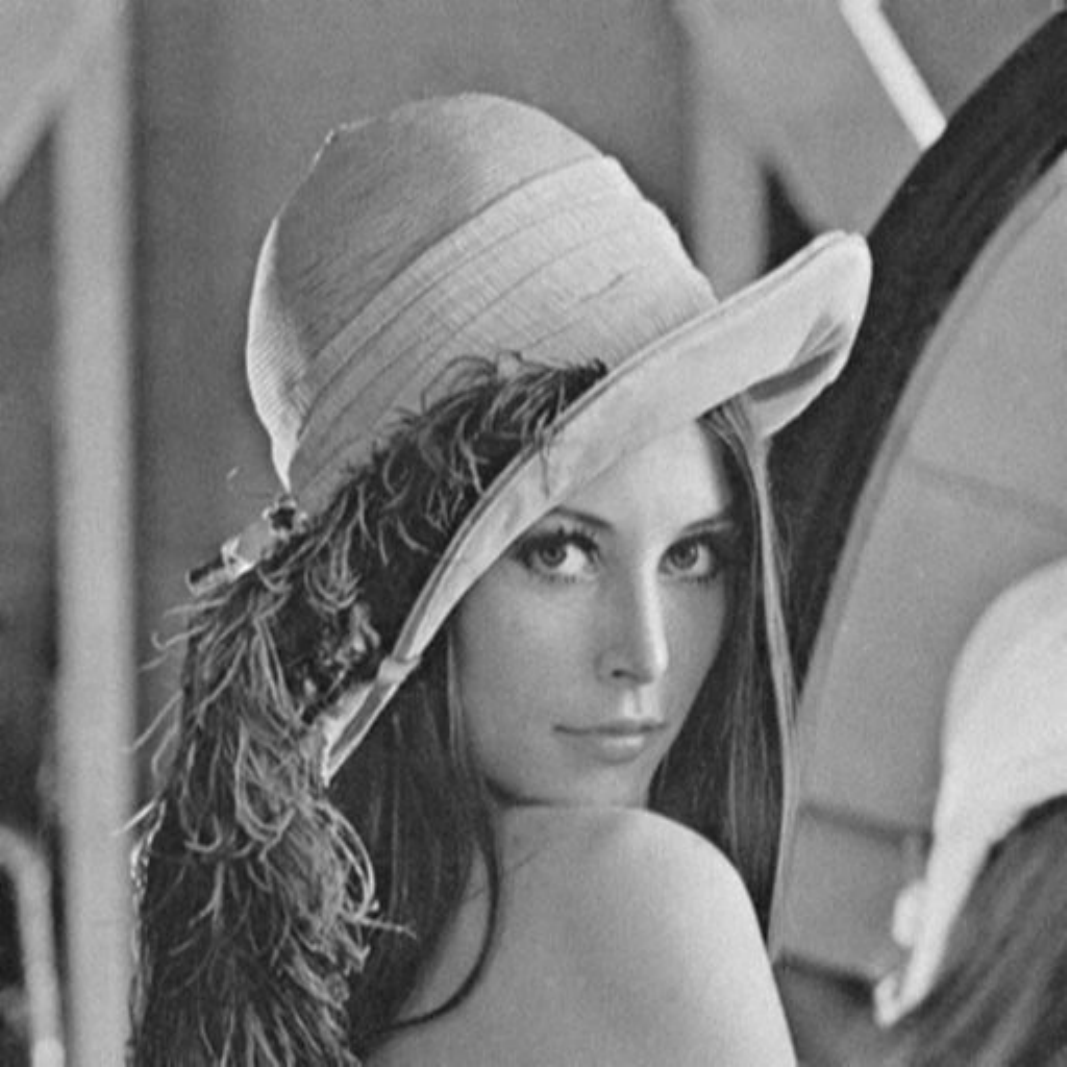}
  \caption{}
\end{subfigure}
\caption{The results for restoring the image \emph{Lena} corrupted by $50\%$ block loss with block size of 4. (a) Missing pattern, (b) Guleryuz\textquoteright{}s method \cite{GuleryuzNonlinearApprox2006} ($29.2$ dB), (c) Takeda et al.\textquoteright{}s method \cite{TakedaKernelRegression2007} ($29.8$ dB), (d) IMAT \cite{MarvastiAminiUnifiedAppr2012} ($14.5$ dB), (e) IMATI \cite{AzghaniSampta2013} ($29.9$ dB), and (f) The proposed method ($85\%$ sparsity rate, $34.1$ dB).}\label{fig:lena}
\end{figure*}

\begin{figure*}[hp]
\centering
\begin{subfigure}[b]{0.35\linewidth}
    \includegraphics[width=\linewidth]{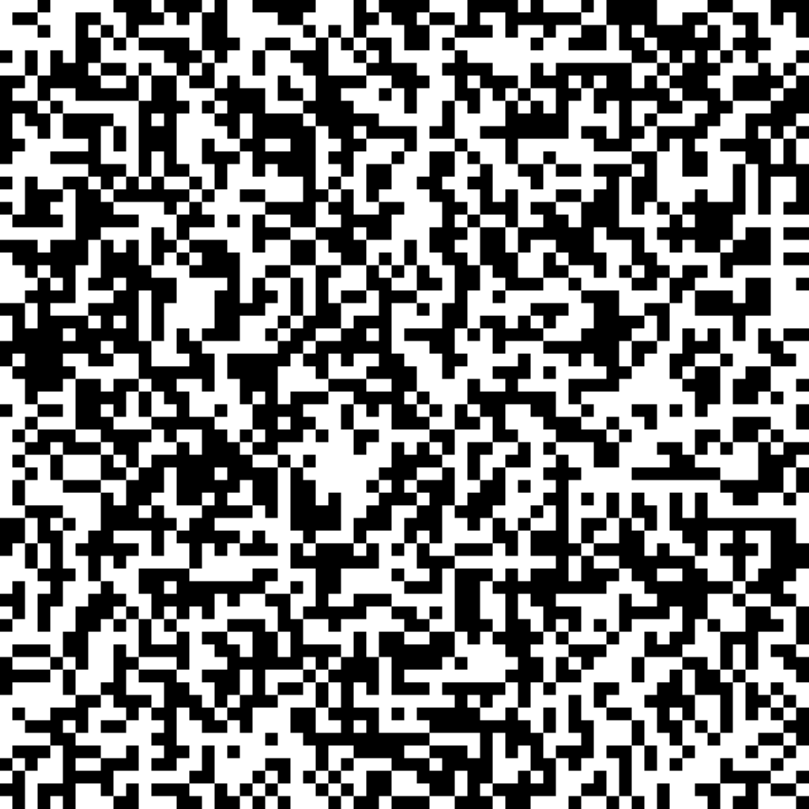}
		  \caption{}
\end{subfigure}
\begin{subfigure}[b]{0.35\linewidth}
    \includegraphics[width=\linewidth]{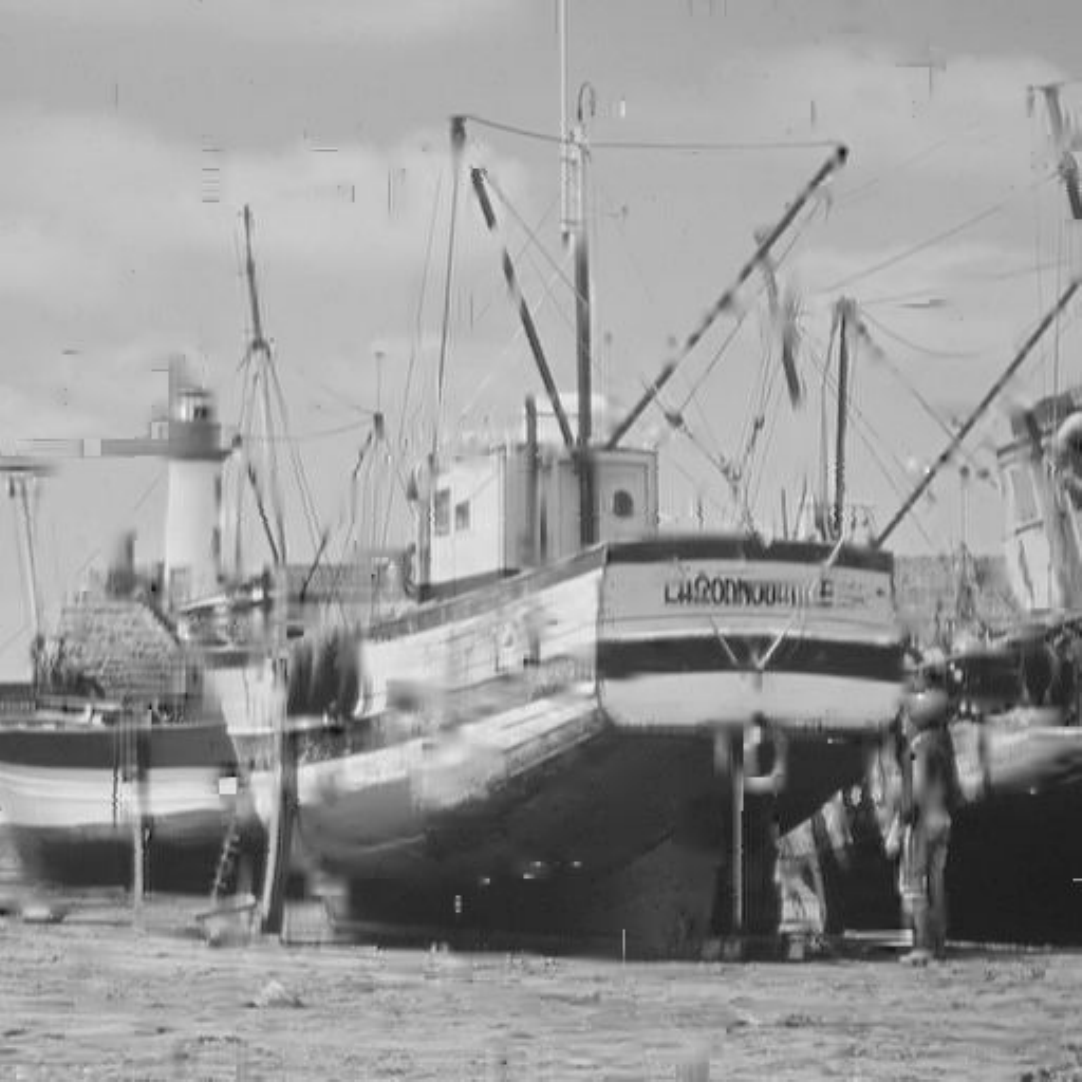}
		  \caption{}
\end{subfigure}
\begin{subfigure}[b]{0.35\linewidth}
    \includegraphics[width=\linewidth]{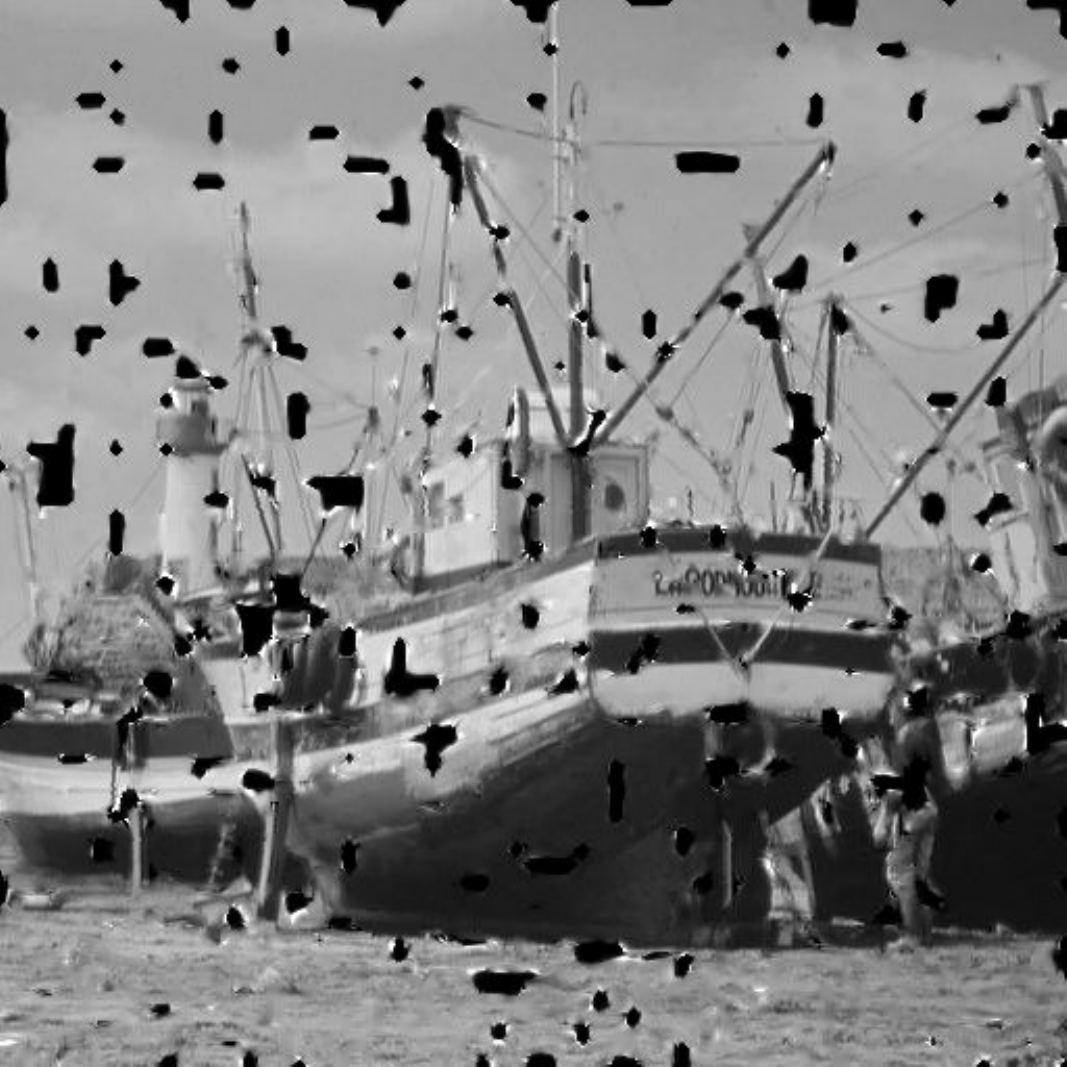}
  \caption{}
\end{subfigure}
\begin{subfigure}[b]{0.35\linewidth}
    \includegraphics[width=\linewidth]{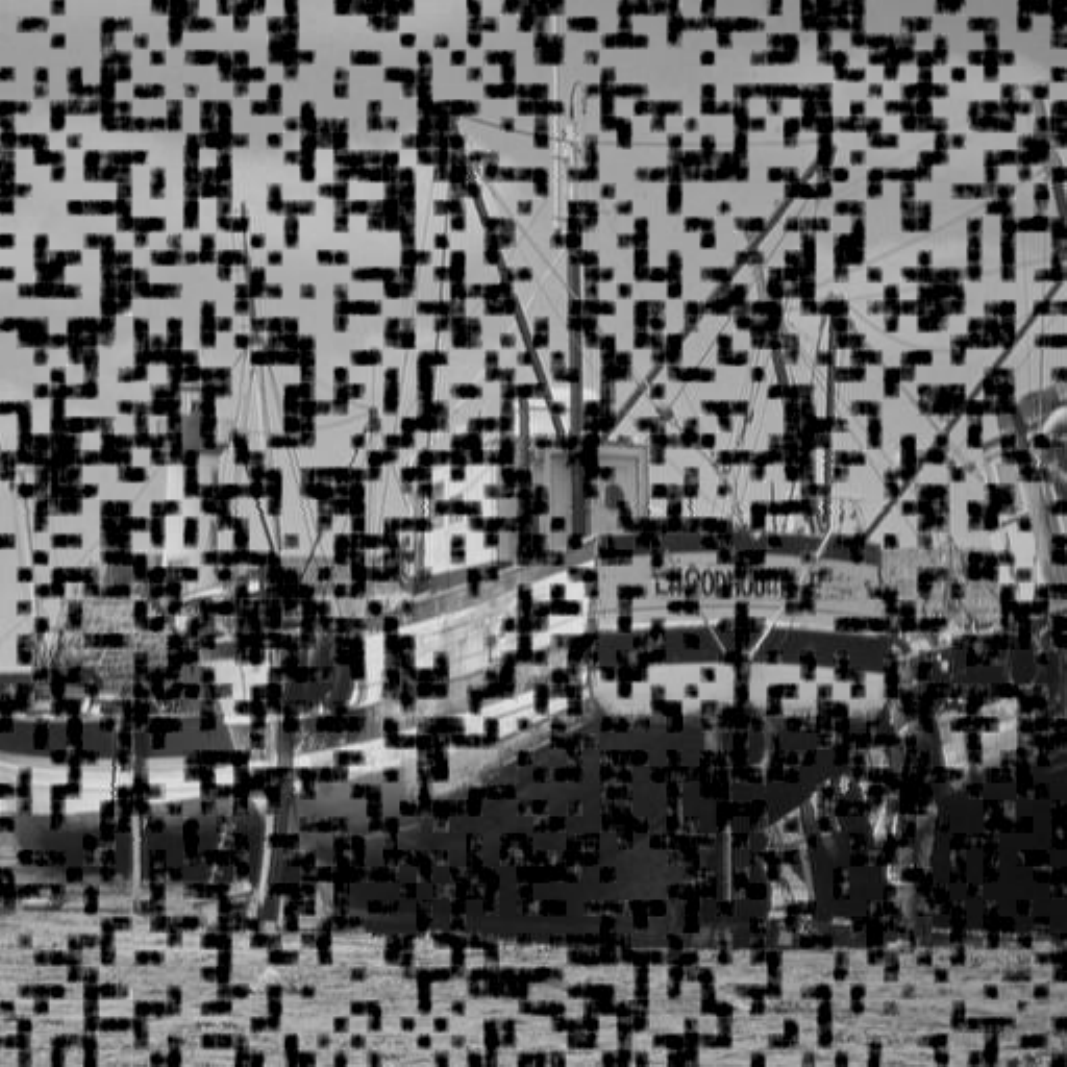}
  \caption{}
\end{subfigure}
\begin{subfigure}[b]{0.35\linewidth}
    \includegraphics[width=\linewidth]{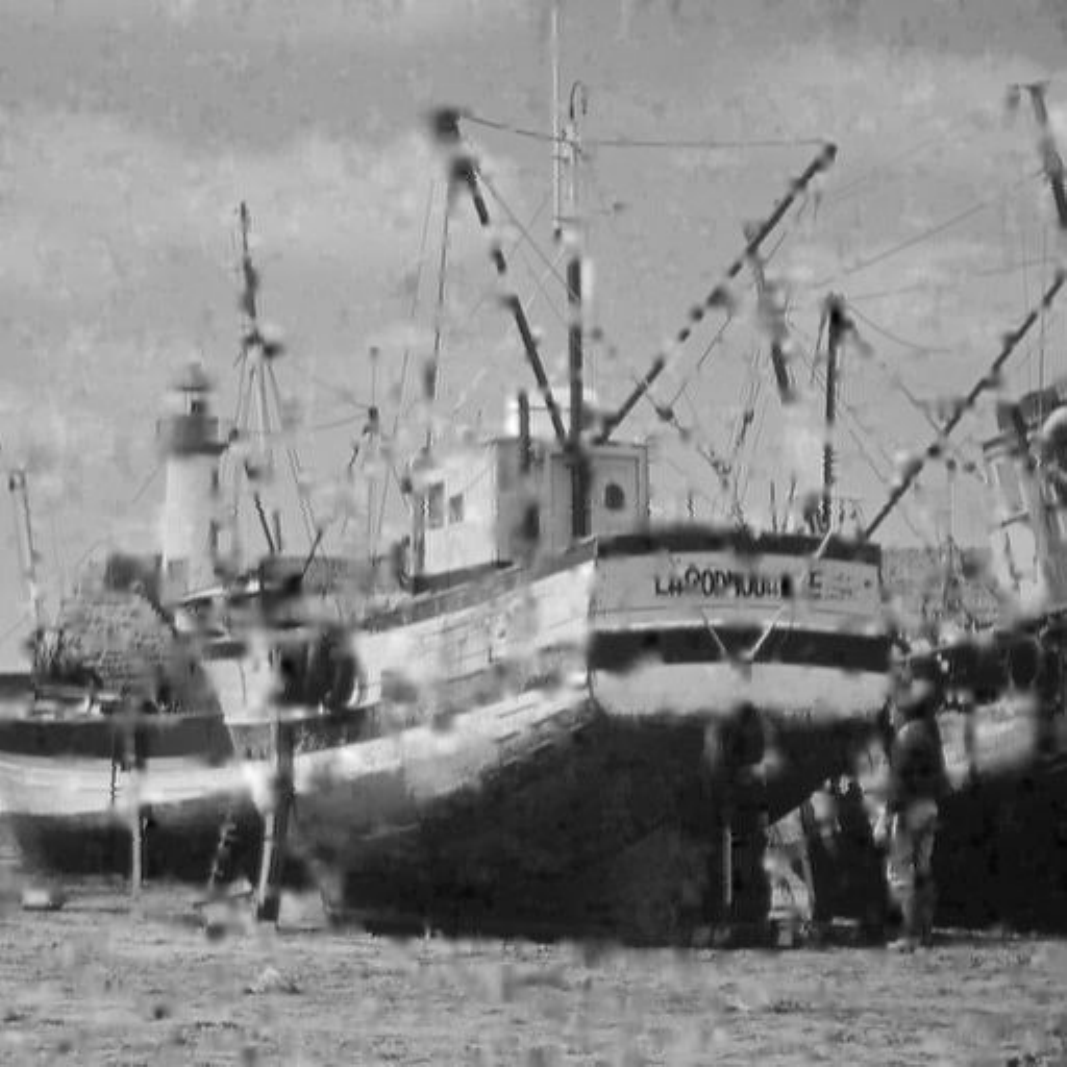}
  \caption{}
\end{subfigure}
\begin{subfigure}[b]{0.35\linewidth}
    \includegraphics[width=\linewidth]{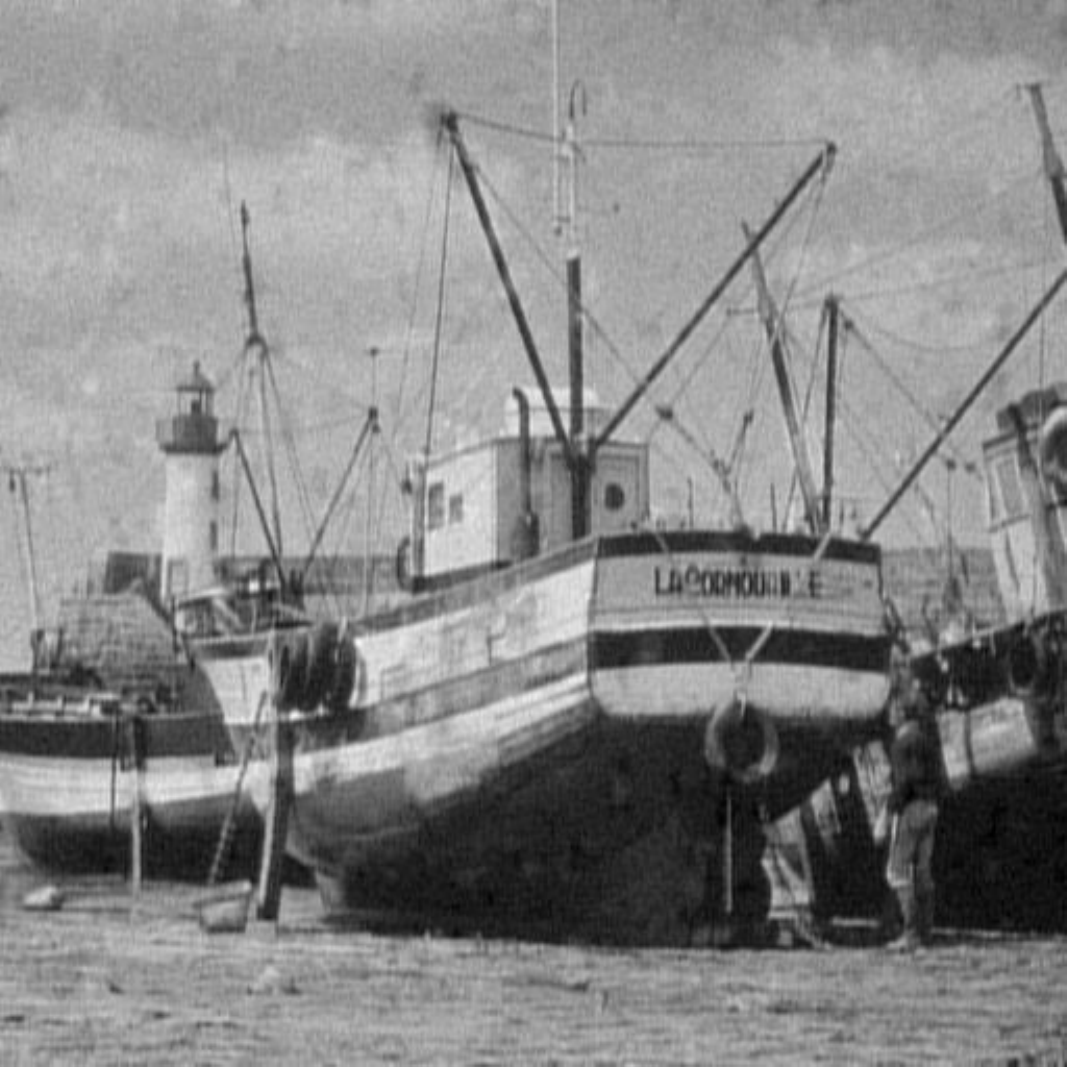}
  \caption{}
\end{subfigure}
\caption{The results for restoring the image \emph{Boat} corrupted by $50\%$ block loss with block size of $8$. (a) Missing pattern, (b) Guleryuz\textquoteright{}s method \cite{GuleryuzNonlinearApprox2006} ($25.2$ dB), (c) Takeda et al.\textquoteright{}s method \cite{TakedaKernelRegression2007} ($16.6$ dB), (d) IMAT \cite{MarvastiAminiUnifiedAppr2012} ($10.4$ dB), (e) IMATI \cite{AzghaniSampta2013} ($24.9$ dB), and (f) The proposed method ($90\%$ sparsity rate, $29.5$ dB).}\label{fig:boat}
\end{figure*}

\begin{figure*}[hp]
\centering
\begin{subfigure}[b]{0.35\linewidth}
    \includegraphics[width=\linewidth]{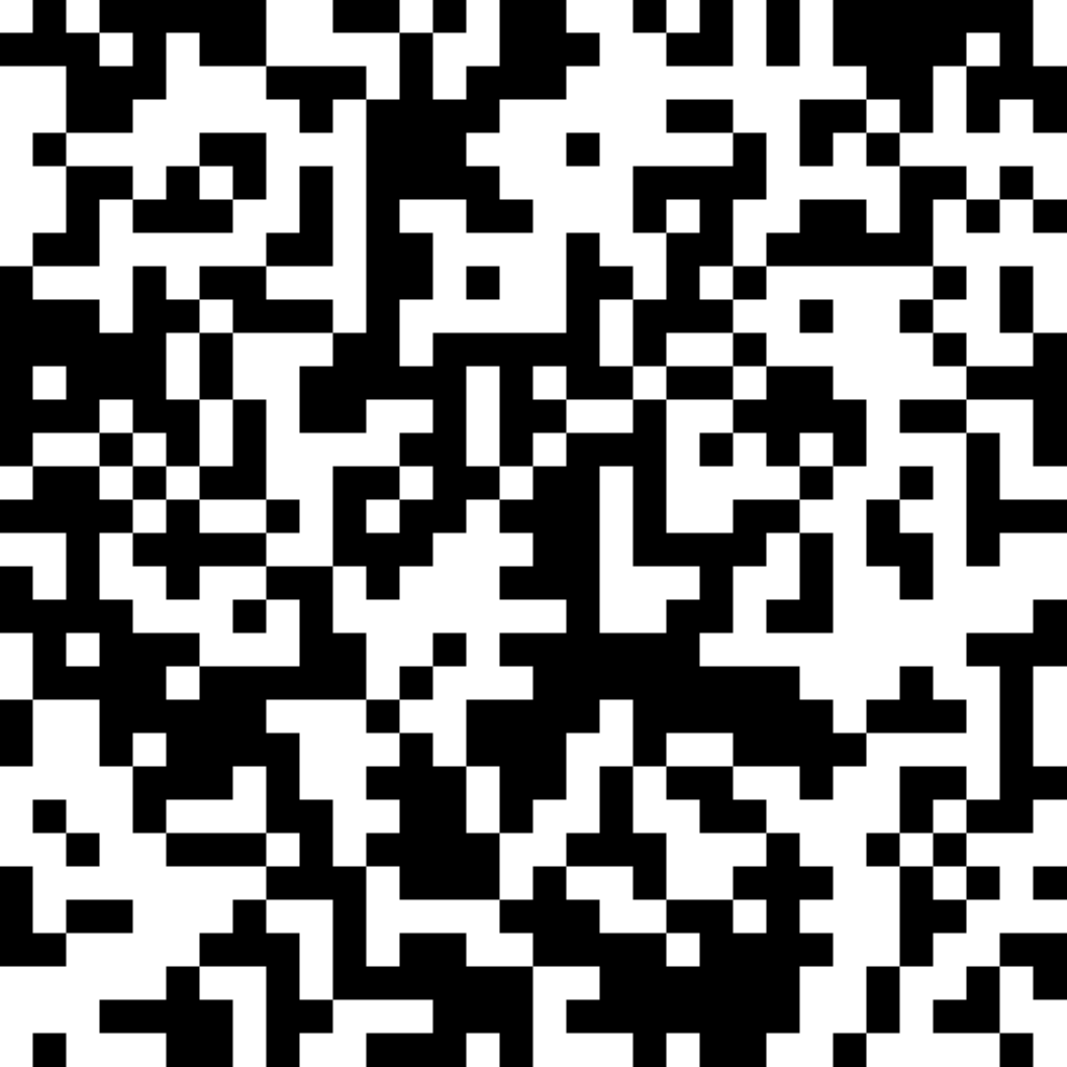}
		  \caption{}
\end{subfigure}
\begin{subfigure}[b]{0.35\linewidth}
    \includegraphics[width=\linewidth]{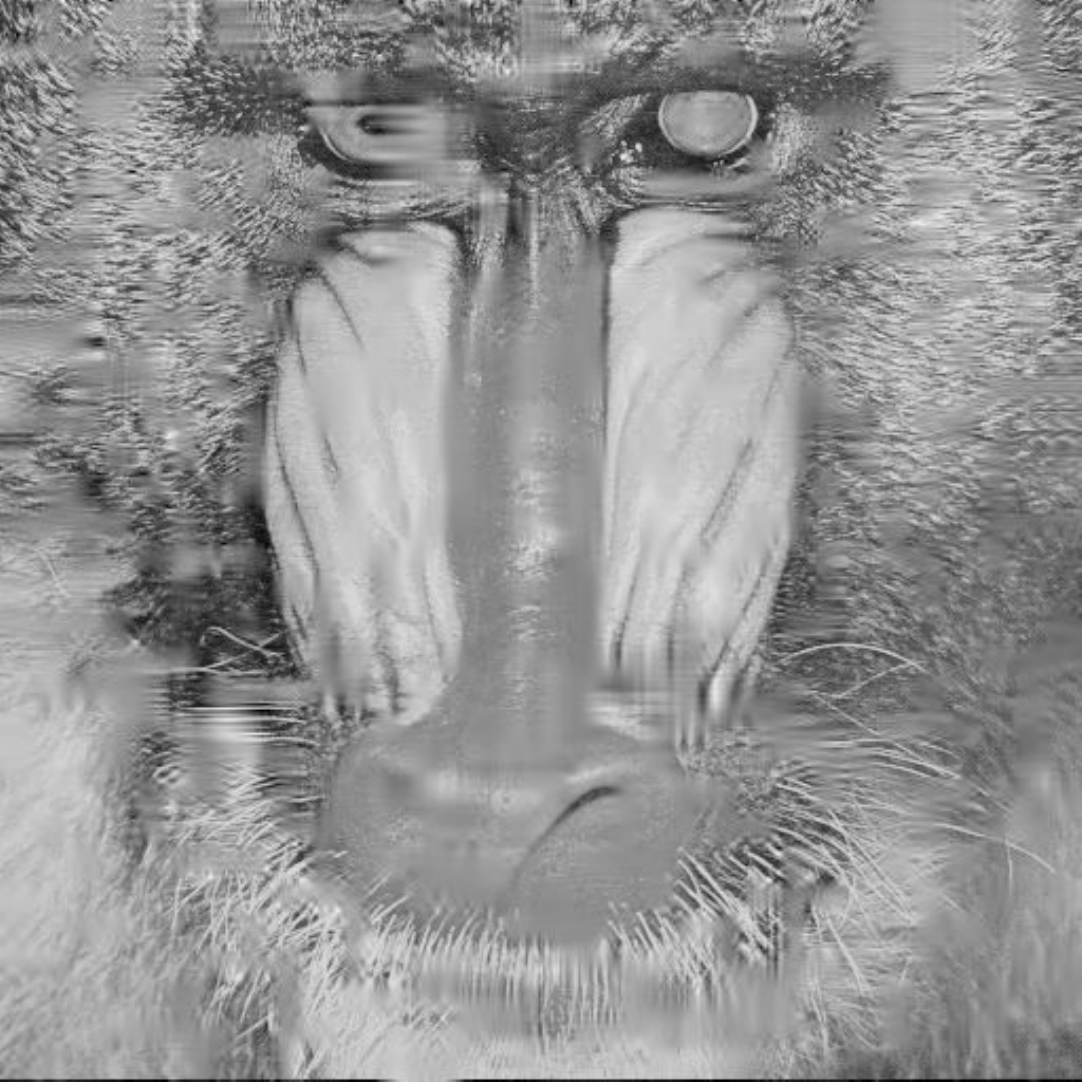}
		  \caption{}
\end{subfigure}
\begin{subfigure}[b]{0.35\linewidth}
    \includegraphics[width=\linewidth]{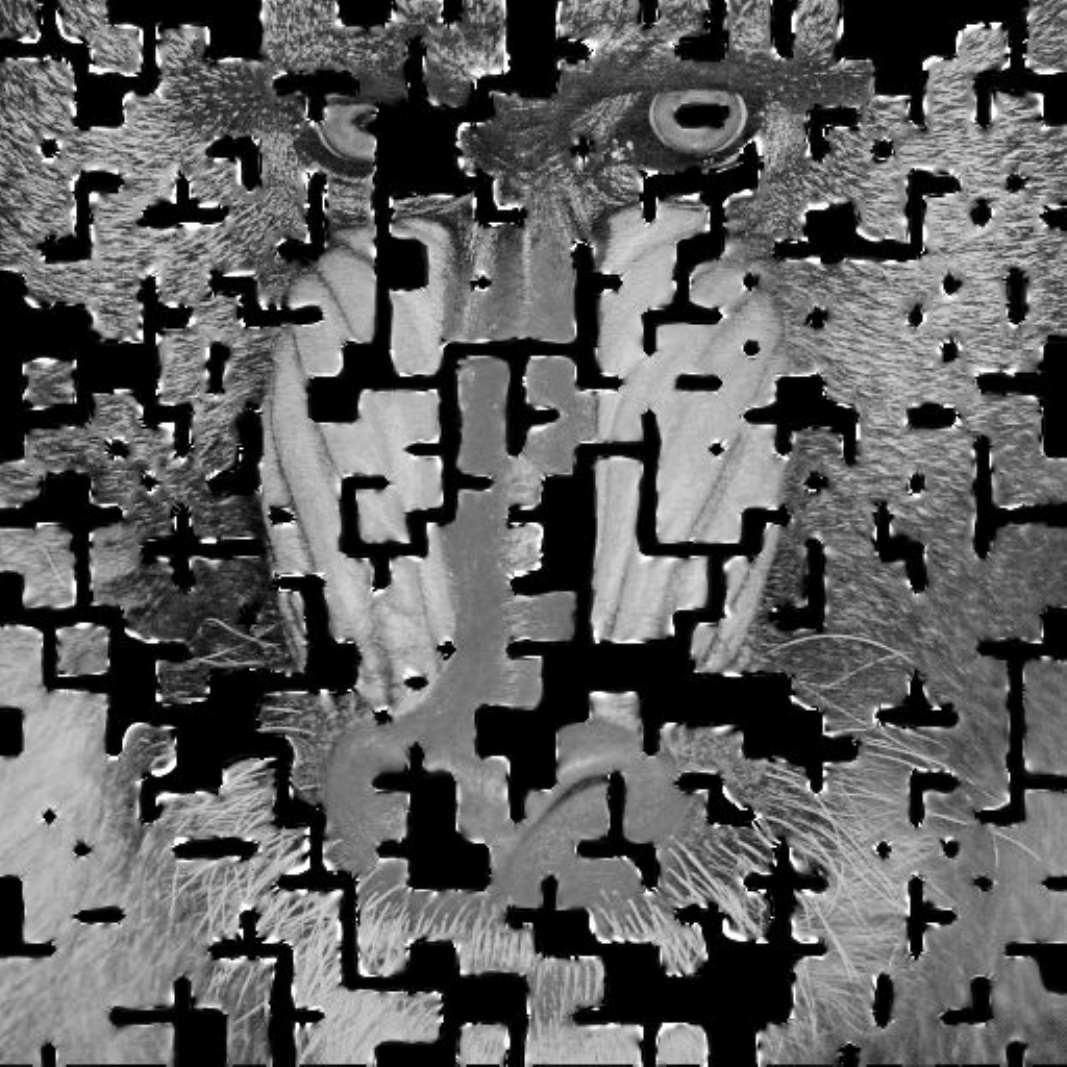}
  \caption{}
\end{subfigure}
\begin{subfigure}[b]{0.35\linewidth}
    \includegraphics[width=\linewidth]{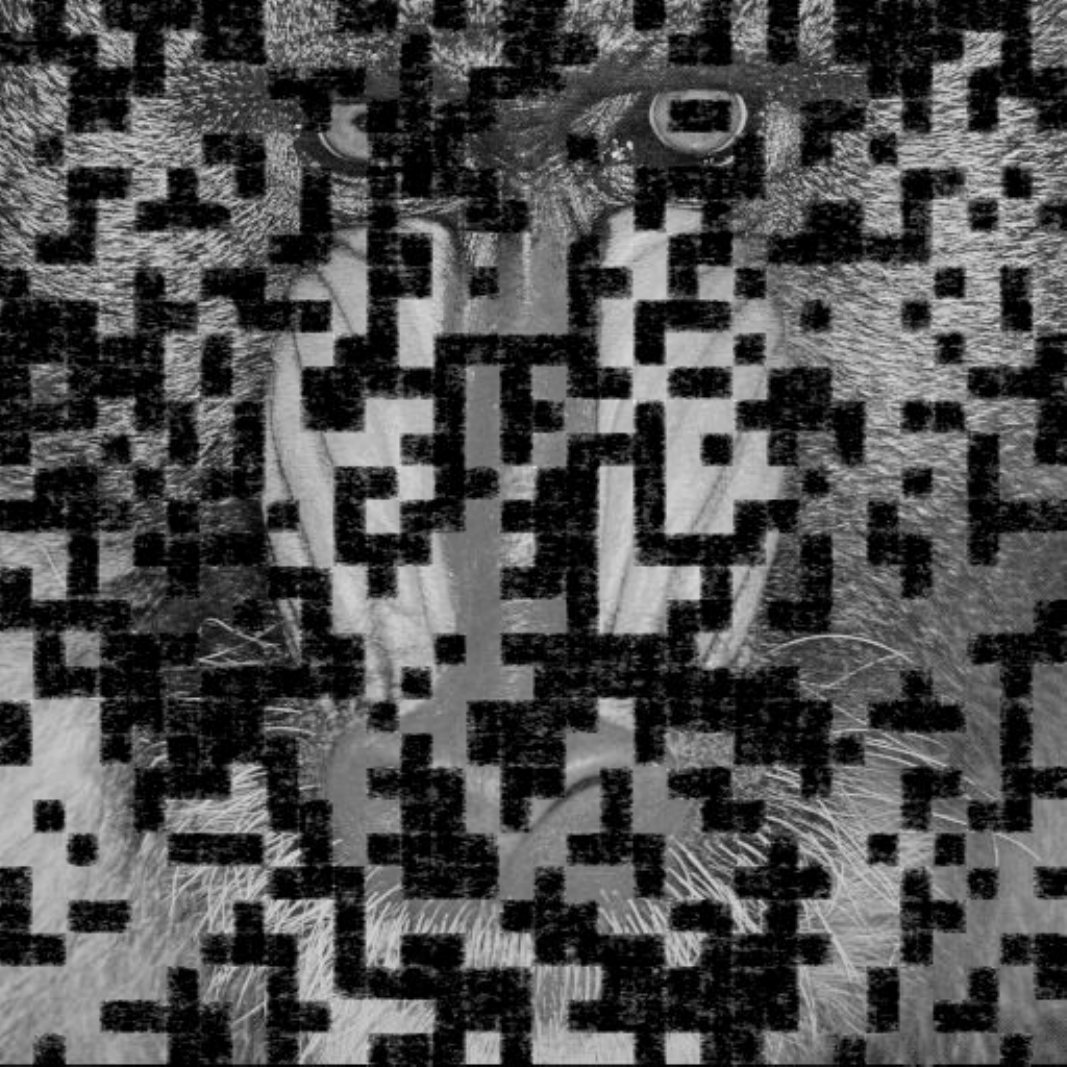}
  \caption{}
\end{subfigure}
\begin{subfigure}[b]{0.35\linewidth}
    \includegraphics[width=\linewidth]{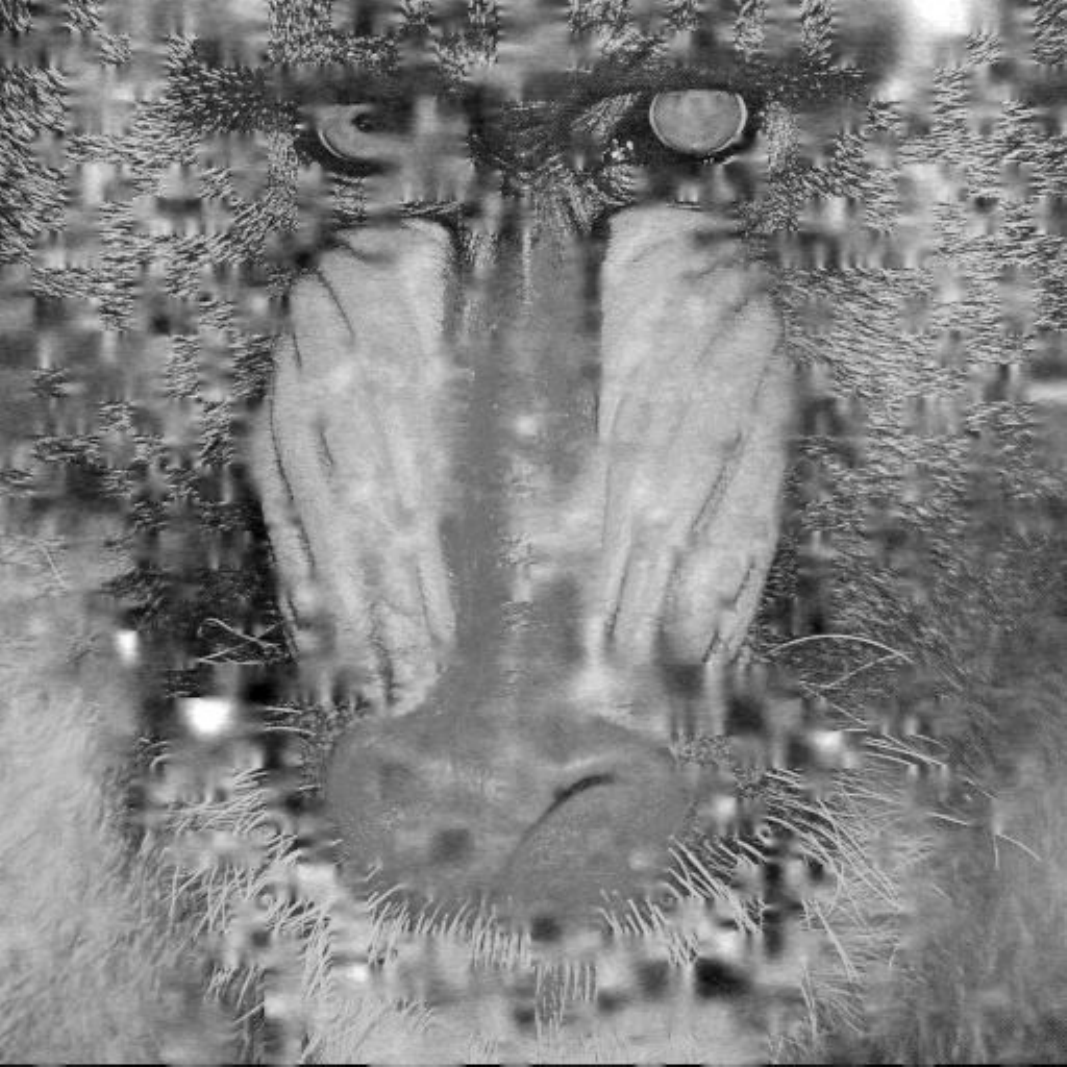}
  \caption{}
\end{subfigure}
\begin{subfigure}[b]{0.35\linewidth}
    \includegraphics[width=\linewidth]{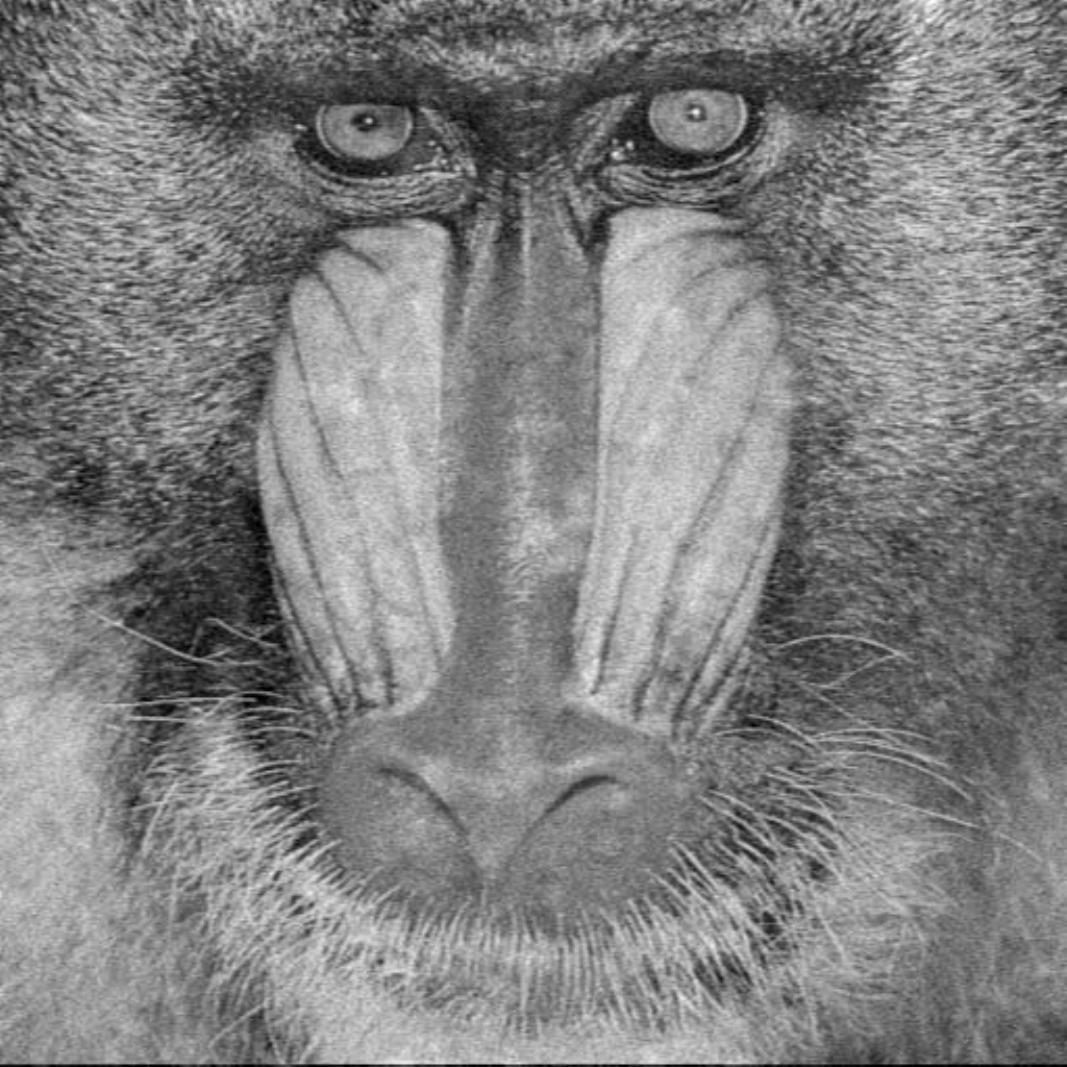}
  \caption{}
\end{subfigure}
\caption{The results for restoring the image \emph{Baboon} corrupted by $50\%$ block loss with block size of $16$. (a) Missing pattern, (b) Guleryuz\textquoteright{}s method \cite{GuleryuzNonlinearApprox2006} ($20.7$ dB), (c) Takeda et al.\textquoteright{}s method \cite{TakedaKernelRegression2007} ($10.8$ dB), (d) IMAT \cite{MarvastiAminiUnifiedAppr2012} ($10.0$ dB), (e) IMATI \cite{AzghaniSampta2013} ($19.7$ dB), and (f) The proposed method ($85\%$ sparsity rate, $23.9$ dB).}\label{fig:baboon}
\end{figure*}

\begin{figure*}[hp]
\centering
\begin{subfigure}[b]{0.45\linewidth}
    \includegraphics[width=\linewidth]{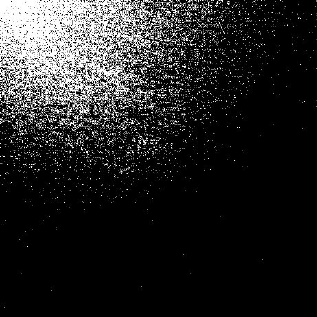}
		  \caption{}
\end{subfigure}
\begin{subfigure}[b]{0.45\linewidth}
    \includegraphics[width=\linewidth]{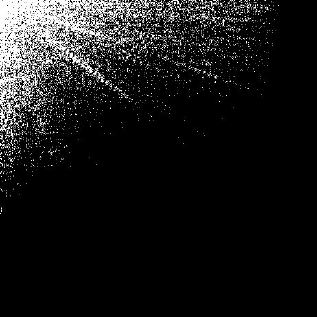}
		  \caption{}
\end{subfigure}
\begin{subfigure}[b]{0.45\linewidth}
    \includegraphics[width=\linewidth]{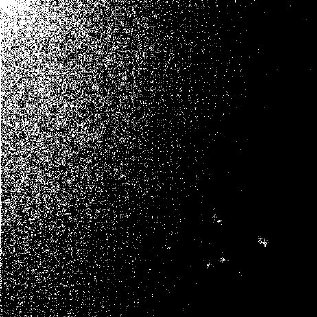}
  \caption{}
\end{subfigure}
\caption{The sparsity patterns which are used in the proposed method for restoring the three images in Figures~\ref{fig:lena} -~\ref{fig:baboon}. (a) Image \emph{Lena} ($85\%$ sparsity rate), (b) Image \emph{Boat} ($90\%$ sparsity rate), and (c) Image \emph{Baboon} ($85\%$ sparsity rate). Sparsity pattern is a binary matrix and the black portion represents zero coefficients.}\label{fig:SPs}
\end{figure*}

The proposed method is also compared with the methods IMAT \cite{MarvastiAminiUnifiedAppr2012} and IMATI \cite{AzghaniSampta2013}, which exploit the image sparsity property, and the algorithms introduced by Guleryuz \cite{GuleryuzNonlinearApprox2006} and Takeda et al. \cite{TakedaKernelRegression2007}. We adopt our technique for stopping the iterations of IMAT and IMATI, since none of them provide a stopping rule. For better comparisons, the AIM interpolator is used in IMATI, which produces better results than the original interpolator. Figure~\ref{fig:block} compares different methods for restoring the three images that are corrupted by block loss with various block sizes. For visual comparison of different methods, we consider three cases of restoring the images \emph{Lena}, \emph{Boat} and \emph{Baboon} that are corrupted by block losses with block sizes of $4$, $8$ and $16$, respectively. Figures ~\ref{fig:lena} -~\ref{fig:baboon} depict the missing patterns and restored images. It can be seen that the proposed method has superior performance in all cases, especially for large block losses. Moreover, Figure \ref{fig:SPs} demonstrates the 2D sparsity patterns that are used in our proposed method for restoring the three images in Figures~\ref{fig:lena} -~\ref{fig:baboon}. It is clear that there is a strong inclination of non-zero coefficients toward low frequency coefficients. This feature has been used to compress the sparsity pattern before embedding it in the sparse image LSBs.

Since in our method the recipient has some knowledge about the transmitted image, the comparison might seem unfair; however, we do the comparisons to justify our thesis that the quality of the transmitted image can be sacrificed for a better missing pixels restoration in the receiver. Certainly, there is a trade-off between the amount of the image degradation, resulted from sparsifying and data embedding, and the capability of the missing pixels restoration. Simulation results verify that by choosing an appropriate sparsity rate, which is possible by well-estimating the missing pattern, we can achieve the best trade-off and, therefore, ensure much higher degree of the overall fidelity to the original image.

\section{CONCLUSION AND FUTURE WORKS}\label{sect:conc}

In this paper, we used side information for image block loss restoration. In the proposed iterative method, the sparsity projector adds a redundancy to the image by discarding the small coefficients of the DCT domain. The resultant sparsity pattern is used as side information in the restoration stage. Two novel features, including a pre-interpolation and a criterion for stopping the iterations, are employed for performance improvement. The method can restore large missing regions at the expense of blurring the transmitted image. Besides, to deal with practical applications, a technique is developed to transmit the side information along with the image. This technique first compresses the sparsity pattern and then embeds its LDPC coded version in the least significant bits of the image pixels, to ensure the error-free transmission of the side information by causing only a small perturbation on the transmitted image. Mathematical analysis is performed to support the efficiency of the introduced techniques. The experimental results verify that the proposed method outperforms the existing algorithms for image block loss restoration.

As mentioned, there is a trade-off between the amount of the image blurring, resulted from sparsifying, and the performance of the block loss restoration. In simulations, for each image and missing pattern, the optimum sparsity rate is specified. However, mathematically determining the optimum sparsity rate has been left for future works.

The proposed method can be employed to efficiently combat the block loss in the compression algorithms. For the transform-based compression methods, like JPEG, we suggest to combine the sparsity projector and the compression function to achieve more effectiveness. We can also use the proposed method for video block loss restoration. Since consecutive frames are roughly similar, we can assign one sparsity pattern for a series of frames and therefore transmit less side information. The set of new zero coefficients should be chosen as the intersection of the corresponding sets of the consecutive frames because, as stated in the paper, unlike the false positive, false negative does not affect the convergence.

\section*{ACKNOWLEDGMENT}

The authors would like to thank Mohammad Mahdi Kamani for helping with some
parts of the simulations.

\bibliographystyle{ieeetr}
\bibliography{Main}
\end{document}